\documentclass[
 12pt, draftclsnofoot, onecolumn
]
{IEEEtran}
\usepackage[utf8]{inputenc}
\usepackage[cmex10]{amsmath}
\usepackage{amssymb,amsthm}

\usepackage[caption=false,font=footnotesize]{subfig}

\usepackage{pgf}
\usepackage{tikz}
\usepackage[utf8]{inputenc}
\usetikzlibrary{arrows,automata}
\usetikzlibrary{positioning}
\usetikzlibrary{decorations.pathmorphing}
\usetikzlibrary{decorations.pathreplacing}
\usetikzlibrary{shapes.geometric}
\usetikzlibrary{fit}
\usetikzlibrary{backgrounds}

\usepackage{color}

\newtheorem{theorem}{Theorem}
\newtheorem{lem}{Lemma}

\theoremstyle{remark}
\newtheorem{rem}{Remark}
\theoremstyle{definition}
\newtheorem{defn}{Definition}
\newtheorem{example}{Example}

\newcommand{\mbf}{\mathbf}
\newcommand{\mbb}{\mathbb}
\newcommand{\mc}{\mathcal}

\newcommand{\e}[1]{^{(#1)}}
\newcommand{\A}[2][1]{\mc A_{s,c}\e{#1}(#2)}

\DeclareMathOperator{\ran}{\mathrm{ran}}
\DeclareMathOperator{\diam}{\mathrm{diam}}

\DeclareMathOperator{\vol}{\mathrm{vol}}

\DeclareMathOperator{\card}{\sharp}

\title{Secure Estimation and\\ Zero-Error Secrecy Capacity}

\author{Moritz~Wiese~\IEEEmembership{Member,~IEEE,} Tobias J. Oechtering~\IEEEmembership{Senior Member,~IEEE,} \\Karl Henrik Johansson~\IEEEmembership{Fellow,~IEEE,} Panos Papadimitratos~\IEEEmembership{Member,~IEEE,}\\ Henrik Sandberg~\IEEEmembership{Member,~IEEE,} Mikael Skoglund~\IEEEmembership{Senior Member,~IEEE}%
\thanks{M.~Wiese, T.~J.~Oechtering and M.~Skoglund are with the Department for Communication Theory, KTH Royal Institute of Technology, Osquldas v\"ag 10, SE-10044 Stockholm, Sweden. K.~H.~Johansson and H.~Sandberg are with the Department for Automatic Control, KTH Royal Institute of Technology, Osquldas v\"ag 10, SE-10044 Stockholm, Sweden. P.~Papadimitratos is with the Networked Systems Security Group, KTH Royal Institute of Technology, Osquldas v\"ag 6, SE-10044 Stockholm, Sweden. e-mail: \{moritzw, oech, kallej, papadim, hsan, skoglund\}@kth.se}
\thanks{This paper was presented in part at the 2016 IEEE International Symposium on Information Theory in Barcelona, Spain and at the 55th IEEE Conference on Decision and Control in Las Vegas, USA.}
}

\begin{document}

\maketitle

\begin{abstract}
  We study the problem of securely estimating the states of an unstable dynamical system subject to nonstochastic disturbances. The estimator obtains all its information through an \textit{uncertain channel} which is subject to nonstochastic disturbances as well, and an eavesdropper obtains a disturbed version of the channel inputs through a second uncertain channel. An encoder observes and block-encodes the states in such a way that, upon sending the generated codeword, the estimator's error is bounded and such that a security criterion is satisfied ensuring that the eavesdropper obtains as little state information as possible. Two security criteria are considered and discussed with the help of a numerical example. A sufficient condition on the \textit{uncertain wiretap channel}, i.e., the pair formed by the uncertain channel from encoder to estimator and the uncertain channel from encoder to eavesdropper, is derived which ensures that a bounded estimation error and security are achieved. This condition is also shown to be necessary for a subclass of uncertain wiretap channels. To formulate the condition, the zero-error secrecy capacity of uncertain wiretap channels is introduced, i.e., the maximal rate at which data can be transmitted from the encoder to the estimator in such a way that the eavesdropper is unable to reconstruct the transmitted data. Lastly, the zero-error secrecy capacity of uncertain wiretap channels is studied.
\end{abstract}

\section{Introduction}

With the increasing deployment and growing importance of cyber-physical systems, the question of their security has recently become a focus of research activity in control theory \cite{CSMCPSSec}. One central vulnerability of networked control or estimation is the communication channel from the system which is to be controlled/estimated to the controller/estimator and possibly the feedback channel. A possible attack on the channels is to actively interfere with transmitted information with the goal of degrading the control or estimation performance. However, if the state of a system is estimated remotely, e.g., in order to decide on the next control action at a remote controller, another possible attack is eavesdropping. An adversary might have the chance to overhear the transmitted information, to make its own state estimate and thus obtain sensitive information. For example, if the system processes health information, leakage of its state might breach privacy. If the system is a production line, knowledge of its state could be valuable information for competitors or for criminals. This paper addresses the question how to protect the transmitted information from such attackers.

We consider an unstable scalar, discrete-time, time-invariant linear system subject to nonstochastic disturbances, where both the initial state and the disturbances are arbitrary elements of a bounded interval. An \textit{estimator} has the goal of estimating the system states in such a way that the supremum over time of the absolute differences between the true state and its estimate is bounded uniformly over all possible system state trajectories. We call this \textit{reliability}. The estimator does not have direct access to the system states. Instead, an \textit{encoder} observes the system state and is linked to the estimator through an \textit{uncertain channel}, where every input is disturbed in a nonstochastic manner and the input and output alphabets are possibly finite. The encoder transforms blocks of state observations into codewords using an \textit{encoding function}, while the estimator applies a \textit{decoding function} for estimating the system states from the channel outputs. Together, the encoding and decoding functions form a \textit{transmission scheme}.

\begin{figure}
	\centering
	\begin{tikzpicture}[kasten/.style={draw, rounded corners, inner sep=4pt, line width = 1pt}, pfeil/.style={->, >=latex, line width = 1pt}, scale=0.8, every node/.style={transform shape}]
    \node[kasten] (plant) {Unstable system};
    
    \node[kasten, below = .5cm of plant] (encoder) {Encoder};

    \node[kasten, below=.5cm of encoder, align=center] (channel) {Uncertain wiretap channel};
    \node[left=.7cm of channel] (aux1) {};
    \node[right=.7cm of channel] (aux2) {};
    
    \node[kasten, below = .7cm of aux1] (estimator) {Estimator};

    \node[kasten, below=.7cm of aux2] (eavesdropper) {Eavesdropper};
    
    \draw[pfeil] (plant) -> (encoder);
    \draw[pfeil] (encoder) -> (channel);
    \draw[pfeil] (channel) -| (estimator);
    \draw[pfeil] (channel) -| (eavesdropper);
 \end{tikzpicture}
	\caption{An unstable system has to be estimated remotely. It obtains state information through an uncertain wiretap channel. The outputs obtained by an eavesdropper at the other channel output need to satisfy an operational security criterion.}\label{fig:genset}
\end{figure}
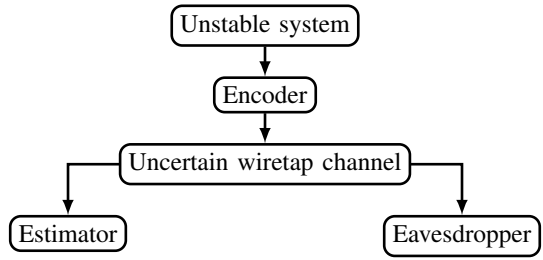

Through another, different, uncertain channel, an adversary called the \textit{eavesdropper} obtains a disturbed version of the encoder's channel input and hence information about the system state. In addition to reliability, our goal is to make the information transmission from the encoder to the estimator secure in such a way that the eavesdropper obtains as little information as possible about the system state, in a sense to be defined. The main question of this paper is under which conditions there exists a transmission scheme such that reliability and security are achieved simultaneously. See Fig.~\ref{fig:genset} for a sketch of the problem setting.

\textit{Contributions:} We introduce the \textit{uncertain wiretap channel}, defined as the pair consisting of the uncertain channel from the encoder to the estimator and the uncertain channel from the encoder to the eavesdropper. We also define the \textit{zero-error secrecy capacity} of the uncertain wiretap channel, which describes the maximal block encoding data rate such that not only the estimator can decode the transmitted message, but at the same time the eavesdropper always has at least two messages among which it cannot distinguish which one was actually transmitted. We show that it either equals zero or the zero-error capacity of the uncertain channel between encoder and estimator. The latter capacity was introduced by Shannon \cite{Shad}. By definition, it is the maximal rate at which, using block encoding, data can be transmitted from the encoder to the estimator through the uncertain channel in such a way that every possible channel output is generated by a unique message. A criterion to distinguish the cases of zero and positive zero-error secrecy capacity can be given in a special case. For the study of the zero-error secrecy capacity of uncertain wiretap channels, we introduce a hypergraph structure on the input alphabet in addition to the graph structure which is applied in the study of the zero-error capacity of uncertain channels and which also goes back to Shannon's original paper \cite{Shad}.

With these information-theoretic tools, we address the main question formulated above. We define two security criteria for secure estimation. The first, called \textit{d-security}, is that there is no possibility for the eavesdropper to process the data it receives in order to obtain a bounded estimation error. The other security criterion is \textit{v-security}, which requires that the volume of the set of system states at a given time which are possible according to the eavesdropper's information should tend to infinity. We identify a sufficient condition which says that reliability and both d- and v-security are achievable if the zero-error secrecy capacity of the uncertain wiretap channel is strictly larger than the logarithm of the coefficient of the unstable system. In the construction of reliable and d- or v-secure transmission schemes, we separate quantization/estimation from channel coding. We also give bounds on the speed of growth of the eavesdropper's estimation error and of the volume of the set of states at a given time which are possible according to the eavesdropper's information. A necessary condition for the simultaneous achievability of reliability, d- and v-security can be given for a subclass of uncertain wiretap channels.

\textit{Related work:} Good overviews over the area of estimation and control under information constraints can be found in the introduction of \cite{MSzero} and in \cite{NFZE}. Matveev and Savkin \cite{MSzero} proved that if the system and channel disturbances are stochastic and the estimator's goal is to obtain an almost surely bounded estimation error, the crucial property of the channel is its Shannon zero-error capacity. This led Nair \cite{Nnst} to introduce a nonstochastic information theory for studying the zero-error capacity of uncertain channels and to consider the problem of estimation and control of linear unstable systems, where the information between sensor and estimator has to be transmitted over an uncertain channel.

There exists a large body of work on information-theoretically secure communication, see \cite{LPS} and \cite{BB}. Stochastic wiretap channels were introduced by \cite{Wy}.
Security in the context of estimation and control has so far mostly meant security against active adversaries, e.g., in \cite{dePTDoS,FTDsec,Gupta2012,PTLP,Teixeira}. To our knowledge, only \cite{LLZ} and \cite{TGP} have combined estimation and security against a passive adversary for an unstable system so far. \cite{LLZ} considers general stochastic disturbances in the system and a stochastic wiretap channel with Gaussian noise and uses a non-operational security criterion based on entropy whose implications are not immediately clear. \cite{TGP} considers a linear system with Gaussian disturbances and Gaussian observation noise, whereas the stochastic wiretap channel randomly and independently deletes input symbols. As a security criterion, \cite{TGP} requires that the eavesdropper's estimation error tends to infinity.

Uncertain channels were introduced by Nair \cite{Nnst}, but were previously considered implicitly in the study of the zero-error capacity of channels with stochastic disturbances as introduced by Shannon \cite{Shad}. The calculation of the zero-error capacity is known as a difficult problem which nowadays is mainly treated in graph theory \cite{KO}.

\textit{Notation:} The cardinality of a finite set $\mc A$ is denoted by $\card\mc A$. If $\card\mc A=1$, we call $\mc A$ a \textit{singleton}. An interval $\mc I$ will also be written $\mc I=[\mc I_{\min},\mc I_{\max}]$. We define the length of $\mc I$ by $\lvert\mc I\rvert$. For two subsets $\mc A,\mc B$ of the real numbers and a scalar $\lambda$, we set $\lambda\mc A+\mc B:=\{\lambda a+b:a\in\mc A,b\in\mc B\}$. A sequence $(a(t))_{t=t_0}^{t_1}$ is denoted by $a(t_0\!:\!t_1)$, where $t_1$ is allowed to equal $\infty$.

\textit{Outline:} In Section \ref{sect:uncch}, uncertain wiretap channels are introduced and the main results concerning their zero-error secrecy capacity are stated. The problem of secure estimation is formulated and the corresponding results are presented in Section \ref{sect:secest}. In Section \ref{sect:quantana}, the quantizers applied in this work are introduced and analyzed. This analysis is used in Section \ref{sect:secestres} for the proof of the results on secure estimation. Section \ref{sect:numex} discusses d- and v-security, including a numerical example. After the conclusion in Section \ref{sect:concl}, Appendix \ref{app:channels} contains the proofs of the results concerning uncertain wiretap channels and some additional discussion, and Appendix \ref{app:quantana} provides the proofs from Section \ref{sect:quantana}.

\section{Uncertain Channels and Uncertain Wiretap Channels}\label{sect:uncch}

Before we can present the model for secure estimation, we need to introduce the model for data communication between the encoder and the receiving parties. This model is the 
uncertain wiretap channel. Since it is new and since some results concerning uncertain wiretap channels are relevant for secure estimation, we devote the complete section to this topic. Our model for secure estimation will be defined in Section \ref{sect:secest}.

\subsubsection{Uncertain Channels}

Let $\mc U,\mc V$ be arbitrary nonempty sets. An \textit{uncertain channel from $\mc U$ to $\mc V$} is a mapping $\mbf U:\mc U\rightarrow2^{\mc V}_*:=2^{\mc V}\setminus\{\varnothing\}$. For any $u\in\mc U$, the set $\mbf U(u)$ is the family of all possible output values of the channel given the input $u$. When transmitting $u$, the output of $\mbf U$ will be exactly one element of $\mbf U(u)$. That $\mbf U(u)\neq\varnothing$ for all $u$ means that every input generates an output. Note that every mapping $\varphi:\mc U\rightarrow\mc V$ can be regarded as an uncertain channel $\Phi:\mc U\rightarrow2^{\mc V}_*$ with singletons as outputs, i.e., $\Phi(u)=\{\varphi(u)\}$. Henceforth, we will not make any notational difference between a mapping and the corresponding uncertain channel. 

\begin{rem}
Note that there are no probabilistic weights on the elements of $\mbf U(u)$. Thus $\mbf U$ models a channel with nonstochastic noise, where $\mbf U(u)$ describes the effect of the noise if the input is $u$.
\end{rem}

We call the set $\ran(\mbf U):=\cup_{u\in\mc U}\mbf U(u)$ the \textit{range of $\mbf U$}. Given two uncertain channels $\mbf U_1:\mc U\rightarrow2^{\mc V}_*$ and $\mbf U_2:\mc V\rightarrow2^{\mc W}_*$, then first applying $\mbf U_1$ and then $\mbf U_2$ leads to a new uncertain channel $\mbf U_2\circ\mbf U_1:\mc U\rightarrow2^{\mc W}_*$ called the \textit{composition of $\mbf U_1$ and $\mbf U_2$}. Formally, we have for any $u\in\mc U$
\[
	(\mbf U_2\circ\mbf U_1)(u):=\mbf U_2(\mbf U_1(u)):=\bigcup_{v\in\mbf U_1(u)}\mbf U_2(v).
\]
Every uncertain channel $\mbf U$ defines a \textit{reverse channel} $\mbf U^{-1}:\ran(\mbf U)\rightarrow2^{\mc U}_*$ by
\[
	\mbf U^{-1}(v)=\{u\in\mc U:v\in\mbf U(u)\}.
\]
Obviously, $\mbf U^{-1}$ again is an uncertain channel. 

\begin{rem}\label{rem:chinj}
    We call $\mbf U^{-1}$ the reverse instead of the inverse because usually, $\card\mbf U^{-1}(\mbf U(u))> 1$. We have $\mbf U^{-1}(\mbf U(u))=\{u\}$ for all $u\in\mc U$ if and only if every output $v\in\ran(\mbf U)$ is generated by exactly one input $u$. If this is the case, we call $\mbf U$ \emph{injective}. If the uncertain channel $\mbf U$ is injective, then $\mbf U^{-1}$ is an ordinary mapping, in the sense that $\mbf U^{-1}(v)$ is a singleton.
\end{rem}

Given uncertain channels $\mbf U_i:\mc U_i\rightarrow2^{\mc V_i}_*\;(1\leq i\leq n)$, their \emph{product} is the channel 
\begin{gather*}
	\mbf U_1\times\cdots\times\mbf U_n:\mc U_1\times\cdots\times\mc U_n\longrightarrow2^{\mc V_1}_*\times\cdots\times 2^{\mc V_n}_*,\\
	(\mbf U_1\times\cdots\times\mbf U_n)(u(1\!:\!n))=\mbf U_1(u_1)\times\cdots\times\mbf U_n(u_n).
\end{gather*}
If $\mbf U_1=\ldots=\mbf U_n=:\mbf U$, we write $\mbf U_1\times\cdots\times\mbf U_n=:\mbf U^n$. The reverse of $\mbf U_1\times\cdots\times\mbf U_n$ is given by $\mbf U_1^{-1}\times\cdots\times\mbf U_n^{-1}$. We write $\mbf U^{-n}$ for the reverse of $\mbf U^n$. 

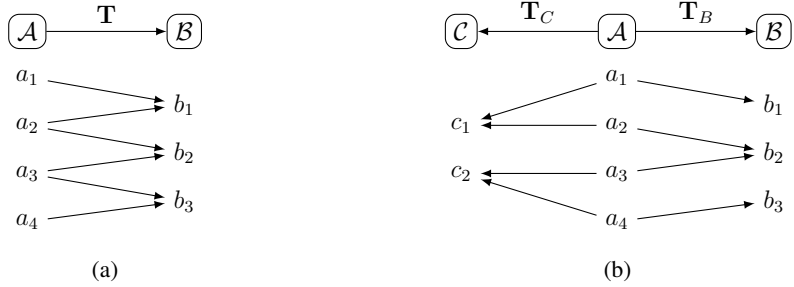
\begin{figure}[!t]
  \centering
  \subfloat[]{
	\begin{tikzpicture}[alphabet/.style={draw, rounded corners}, pfeil/.style={->, >=latex}, scale=0.8, every node/.style={transform shape}]
		\node[alphabet] (Alice) {$\mc A$};
		\node[alphabet, right = 2cm of Alice] (Bob) {$\mc B$};
		
		\node[below = .2cm of Alice] (a1) {$a_1$};
		\node[below = 1cm of Alice] (a2) {$a_2$};
		\node[below = 1.8cm of Alice] (a3) {$a_3$};
		\node[below = 2.6cm of Alice] (a4) {$a_4$};
				
		\node[below = .6cm of Bob] (b1) {$b_1$};
		\node[below = 1.4cm of Bob] (b2) {$b_2$};
		\node[below = 2.2 of Bob] (b3) {$b_3$};
		
		\draw[pfeil] (Alice) -> (Bob) node[midway, above] {$\mbf T$};
		
		\draw[pfeil] (a1) -- (b1);
		\draw[pfeil] (a2) -- (b1);
		\draw[pfeil] (a2) -- (b2);
		\draw[pfeil] (a3) -- (b2);
		\draw[pfeil] (a3) -- (b3);
		\draw[pfeil] (a4) -- (b3);
	\end{tikzpicture}
  }
  \hfil
  \subfloat[]{
  \begin{tikzpicture}[alphabet/.style={draw, rounded corners}, pfeil/.style={->, >=latex}, scale=0.8, every node/.style={transform shape}]
		\node[alphabet] (Eve) {$\mc C$};
		\node[alphabet, right = 2cm of Eve] (Alice) {$\mc A$};
		\node[alphabet, right = 2cm of Alice] (Bob) {$\mc B$};
		
		\node[below = .2cm of Alice] (a1) {$a_1$};
		\node[below = 1cm of Alice] (a2) {$a_2$};
		\node[below = 1.8cm of Alice] (a3) {$a_3$};
		\node[below = 2.6cm of Alice] (a4) {$a_4$};
		
		\node[below = 1cm of Eve] (c1) {$c_1$};
		\node[below = 1.8cm of Eve] (c2) {$c_2$};
		
		\node[below = .6cm of Bob] (b1) {$b_1$};
		\node[below = 1.4cm of Bob] (b2) {$b_2$};
		\node[below = 2.2cm of Bob] (b3) {$b_3$};
		
		\draw[pfeil] (Alice) -> (Bob) node[midway, above] {$\mbf T_B$};
		\draw[pfeil] (Alice) -> (Eve) node[midway, above] {$\mbf T_C$};
		
		\draw[pfeil] (a1) -> (b1);
		\draw[pfeil] (a1) -> (c1);
		\draw[pfeil] (a2) -> (b2);
		\draw[pfeil] (a2) -> (c1);
		\draw[pfeil] (a3) -> (b2);
		\draw[pfeil] (a3) -> (c2);
		\draw[pfeil] (a4) -> (b3);
		\draw[pfeil] (a4) -> (c2);
		
	\end{tikzpicture}
	}
    \caption{(a) An uncertain channel $\mbf T$. If one sets $\mbf F(0)=\{a_1\},\mbf F(1)=\{a_3\}$, then $\mbf F$ is a zero-error 2-code for $\mbf T$. (b) An uncertain wiretap channel $(\mbf T_B,\mbf T_C)$. The uncertain channel $\mbf F:\{0,1,2\}\rightarrow2^{\mc A}_*$ defined by $\mbf F(0)=\{a_1\},\mbf F(1)=\{a_2,a_3\},\mbf F(2)=\{a_4\}$ is a zero-error wiretap 3-code for $(\mbf T_B,\mbf T_C)$.}\label{fig:uncch-code}
\end{figure}

\subsubsection{Zero-Error Codes}

An \textit{$M$-code} on an alphabet $\mc A$ is a collection $\{\mbf F(m):0\leq m\leq M-1\}$ of nonempty and mutually disjoint subsets of $\mc A$. This is equivalent to an uncertain channel $\mbf F:\{0,\ldots,M-1\}\rightarrow2^{\mc A}_*$ with disjoint output sets, so we will often denote such a code just by $\mbf F$. The elements of $\ran(\mbf F)$ are called \textit{codewords}. If $\card\mbf F(m)=1$ for all $0\leq m\leq M-1$, then we call $\mbf F$ a \textit{singleton code}. Zero-error codes which are not singleton codes are introduced here for the first time. 

Let $\mbf T:\mc A\rightarrow2^{\mc B}_*$ be an uncertain channel over which data are to be transmitted. A nonstochastic $M$-code $\mbf F$ on $\mc A$ is called a \textit{zero-error $M$-code for $\mbf T$} if  for any $m,m'\in\{0,\ldots,M-1\}$ with $m\neq m'$
\begin{equation}\label{eq:reliability}
	\mbf T(\mbf F(m))\cap\mbf T(\mbf F(m'))=\varnothing.
\end{equation}
Thus every possible channel output $y\in\ran(\mbf T\circ\mbf F)$ can be associated to a unique message $m$. In other words, the channel $\mbf T\circ\mbf F$ is injective, or equivalently, $\mbf F^{-1}\circ\mbf T^{-1}$ is an ordinary mapping associating to each output $y$ the message $\mbf F^{-1}(\mbf T^{-1}(y))$ by which it was generated (cf.~Remark \ref{rem:chinj}). 
See Fig.~\ref{fig:uncch-code}(a) for an illustration.

\subsubsection{Uncertain Wiretap Channels and Zero-Error Wiretap Codes}

Given an additional finite alphabet $\mc C$, an \textit{uncertain wiretap channel} is a pair of uncertain channels $(\mbf T_B:\mc A\rightarrow2^{\mc B}_*,\mbf T_C:\mc A\rightarrow2^{\mc C}_*)$. The interpretation is that the outputs of channel $\mbf T_B$ are received by an intended receiver, whereas the outputs of $\mbf T_C$ are obtained by an eavesdropper who should not be able to learn the data transmitted over $\mbf T_B$.

An $M$-code $\mbf F$ is called a \textit{zero-error wiretap $M$-code for $(\mbf T_B,\mbf T_C)$} if it is a zero-error $M$-code for $\mbf T_B$ and additionally  
\begin{equation}\label{eq:security}
	\card\mbf F^{-1}(\mbf T_C^{-1}(c))\geq 2
\end{equation}
for every $c\in\ran(\mbf T_C\circ\mbf F)$. Thus every output $c\in\ran(\mbf T_C\circ\mbf F)$ can be generated by at least two messages. Due to the lack of further information like stochastic weights on the messages conditional on the output, the eavesdropper is unable to distinguish these messages. See Fig.\ \ref{fig:uncch-code}(b) for an example.

\subsubsection{Zero-Error Capacity and Zero-Error Secrecy Capacity}

Given an uncertain channel $\mbf T:\mc A\rightarrow 2^{\mc B}_*$, an $M$-code $\mbf F$ on $\mc A^n$ is called a \textit{zero-error $(n,M)$-code for $\mbf T$} if it is a zero-error $M$-code for $\mbf T^n$. We call $n$ the \textit{blocklength} of $\mbf F$. We set $N_{\mbf T}(n)$ to be the maximal $M$ such that there exists a zero-error $(M,n)$-code for $\mbf T$ and define the \textit{zero-error capacity of $\mbf T$} by
\begin{equation}\label{eq:C0}
	C_0(\mbf T):=\sup_n\frac{\log N_{\mbf T}(n)}{n}.
\end{equation}
Due to the superadditivity of the sequence $\log N_{\mbf T}(0\!:\!\infty)$ and  Fekete's lemma \cite{Fek}, see also \cite[Lemma 11.2]{CK}, the supremum on the right-hand side of \eqref{eq:C0} can be replaced by a $\lim_{n\rightarrow\infty}$. Thus $C_0(\mbf T)$ is the asymptotically largest exponential rate at which the number of messages which can be transmitted through $\mbf T$ free of error grows in the blocklength.

Given an uncertain wiretap channel $(\mbf T_B,\mbf T_C)$, a zero-error $(n,M)$-code $\mbf F$ for $\mbf T_B$ is called a \textit{zero-error wiretap $(n,M)$-code for $(\mbf T_B,\mbf T_C)$}  if it is a zero-error wiretap $M$-code for $(\mbf T_B^n,\mbf T_C^n)$. We define $N_{(\mbf T_B,\mbf T_C)}(n)$ to be the maximal $M$ such that there exists a zero-error wiretap $(M,n)$-code for $(\mbf T_B,\mbf T_C)$. If no zero-error wiretap $(n,M)$-code exists, we set $N_{(\mbf T_B,\mbf T_C)}(n)=1$. The \textit{zero-error secrecy capacity of $(\mbf T_B,\mbf T_C)$} is defined as
\begin{align}
	C_0(\mbf T_B,\mbf T_C):=\sup_n\frac{\log N_{(\mbf T_B,\mbf T_C)}(n)}{n}.\label{eq:C0WT}
\end{align}
Again by superadditivity and Fekete's lemma \cite{Fek, CK}, the supremum in \eqref{eq:C0WT} can be replaced by a limit. Obviously, $C_0(\mbf T_B,\mbf T_C)\leq C_0(\mbf T_B)$.

\subsubsection{Capacity Results}

The zero-error capacity of general uncertain channels is unknown, only a few special cases have been solved so far \cite{KO}. However, it is possible to relate the zero-error secrecy capacity of an uncertain wiretap channel $(\mbf T_B,\mbf T_C)$ to the zero-error capacity of $\mbf T_B$ in a surprisingly simple way. 

\begin{theorem}\label{thm:zescap}
	The zero-error secrecy capacity of an uncertain wiretap channel $(\mbf T_B,\mbf T_C)$ either equals 0 or $C_0(\mbf T_B)$.
\end{theorem}

The proof of this result can be found in Appendix \ref{app:channels}. The simple observation behind the proof is that the possibility of sending one bit securely over $(\mbf T_B,\mbf T_C)$ as a prefix to an arbitrary zero-error code $\mbf F$ for $\mbf T_B$ generates a zero-error wiretap code for $(\mbf T_B,\mbf T_C)$ whose rate is approximately the same as that of $\mbf F$.

What is missing in Theorem \ref{thm:zescap} is a necessary and sufficient criterion for the zero-error secrey capacity to be positive. We can give one in the case that $\mbf T_B$ is injective and the input alphabet is finite. 

\begin{theorem}\label{thm:injch}
	Let $(\mbf T_B,\mbf T_C)$ be an uncertain wiretap channel with finite input alphabet $\mc A$ such that $\mbf T_B$ is injective. Then $C_0(\mbf T_B,\mbf T_C)=0$ if and only if $N_{(\mbf T_B,\mbf T_C)}(1)=1$. If $C_0(\mbf T_B,\mbf T_C)>0$, then $C_0(\mbf T_B,\mbf T_C)=\log(\card\mc A)$. 
\end{theorem}

The proof can be found in Appendix \ref{app:channels}. Theorem \ref{thm:injch} gives a characterization of the positivity of the zero-error secrecy capacity if $\mbf T_B$ is injective which only involves $(\mbf T_B,\mbf T_C)$ at blocklength 1. Its proof also contains a simple procedure for finding $N_{(\mbf T_B,\mbf T_C)}(1)$. If $\mbf T_B$ is not injective, finding $N_{(\mbf T_B,\mbf T_C)}(1)$ is harder, but can be done by brute-force search for reasonably sized alphabets. More importantly, if $\mbf T_B$ is not injective, it is possible that $N_{(\mbf T_B,\mbf T_C)}(1)=1$ and $C_0(\mbf T_B,\mbf T_C)>0$, see Example \ref{ex:superact} in Appendix \ref{app:channels}. For general uncertain wiretap channels, one can use the procedure from the proof of Theorem \ref{thm:injch} to reduce a zero-error code for $\mbf T_B$ to a zero-error wiretap code. However, the code thus generated might have rate 0 although $C_0(\mbf T_B,\mbf T_C)>0$. The question when $C_0(\mbf T_B,\mbf T_C)>0$ for general uncertain wiretap channels seems to be a hard problem and has to be left open for now. Further discussion of zero-error secrecy capacity is included in Appendix \ref{app:channels}.

\subsubsection{Degree of Eavesdropper Ignorance}\label{subsubsect:detmeas}

In order to measure the achieved degree of security in greater detail, we introduce the number of messages that can generate a given eavesdropper output as an additional parameter. We call a zero-error wiretap $(n,M)$-code a zero-error wiretap $(n,M,\gamma)$-code if for every $c(1\!:\!n)\in\ran(\mbf T_C^n\circ\mbf F)$, 
\begin{align}
  \card\mbf F^{-1}(\mbf T_C^{-n}(c(1\!:\!n)))\geq\gamma.\label{eq:sharp}
\end{align}
Clearly, $M\geq\gamma\geq2$. This parameter can be interpreted as a measure of the minimal eavesdropper's confusion about the transmitted message guaranteed by the $(n,M,\gamma)$-code. It will be important in the analysis of one of the security criteria we apply for secure estimation.

\section{Secure Estimation Over Uncertain Channels}\label{sect:secest}

\subsection{The Model}

Let $\mc I_0$ be a closed real interval and let $\Omega\geq 0$ and $\lambda>1$ be real numbers such that $\lvert\mc I_0\rvert+\Omega>0$. We then consider the real-valued time-invariant unstable linear system 
\begin{subequations}\label{eq:dynsyst}
\begin{align}
	x(t+1)&=\lambda x(t)+w(t),\label{eq:dynrec}\\
	x(0)&\in\mc I_0.\label{eq:initstate}
\end{align}
\end{subequations}
The initial state $x(0)$ can assume any value in $\mc I_0$ and is not known before its observation. The noise sequence $w(0:\infty)$ can be any sequence in $[-\Omega/2,\Omega/2]^\infty$. We call $x(t)$ the \textit{system state at time $t$}. The system states are directly observable. Due to $\lvert\mc I_0\rvert + \Omega>0$, the system suffers from nontrivial disturbances in the initial state or in the evolution. The set of possible system trajectories $x(0:t)$ until time $t$ is denoted by $\mc X_{0:t}$.

Assume that an entity called the \textit{encoder} is located at the system output and at time $t$ records the corresponding system state $x(t)$. At every system time step, it has the possibility of using an uncertain wiretap channel $(\mbf T_B:\mc A\rightarrow2^{\mc B}_*,\mbf T_C:\mc A\rightarrow2^{\mc C}_*)$ exactly once, i.e., the system \eqref{eq:dynsyst} and the channel are synchronous. At the output of $\mbf T_B$, an estimator has the task of obtaining reliable estimates of the system states. An eavesdropper has access to the outputs of $\mbf T_C$ which should  satisfy a security criterion. 

At time $t$, the encoder only knows $x(0\!:\!t)$ and the system dynamics \eqref{eq:dynsyst}, i.e., it has no acausal knowledge of future states. The estimator and the eavesdropper know the system dynamics \eqref{eq:dynsyst}, but the only information about the actual system states they have is what they receive from the encoder through $\mbf T_B$ and $\mbf T_C$, respectively. The eavesdropper also knows the transmission protocol applied by encoder and estimator. 

The encoder also has knowledge of the complete uncertain wiretap channel, in particular the characteristics of the uncertain channel to the eavesdropper. This knowledge can be justified by assuming that the eavesdropper is part of the communication network without access rights for the system state, e.g., an ``honest but curious'' node in the home network. Uncertain wiretap channels can also be regarded as models of stochastic wiretap channels where the transition probabilities are unknown. In the other direction, there exist information-theoretic techniques for wiretap channels which do not require precise knowledge about the channel to the eavesdropper, but the case with eavesdropper channel knowledge serves as a building block and as a benchmark \cite{BL, AVWCI}. 

The allowed protocols are defined next.

\begin{defn}\label{def:transmscheme}
	A \textit{transmission scheme} consists of a positive integer $n$ called the \textit{blocklength} of the transmission scheme together with a sequence of pairs $(f_k,\varphi_k)_{k=0}^\infty$. Setting $\tau_k:=kn+1$ and $t_k:=(k+1)n$, for every $k\geq 0$
	\begin{itemize}
		\item the $k$-th \textit{encoding function} $f_k:\mc X_{0:\tau_k-1}\rightarrow 2^{\mc A^{n}}_*$ is an uncertain channel, 
		\item the first \textit{decoding function} $\varphi_1:\mc B^n\rightarrow\mbb R$ is an ordinary mapping,
		\item for $k\geq 2$, the $k$-th \textit{decoding function} $\varphi_k:\mc B^{t_k}\rightarrow\mbb R^{n}$ is an ordinary mapping.
	\end{itemize}
\end{defn}

\begin{figure}
	\centering
	\begin{tikzpicture}[kasten/.style={draw, rounded corners, inner sep=4pt}, pfeil/.style={->, >=latex}, scale=0.8, every node/.style={transform shape}]
    \node (noise) {$w(\tau_{k-1}\!:\!\tau_k-1)$};
    \node[kasten, below=.5cm of noise] (plant) {$x(\tau_{k-1}\!:\!\tau_k-1)$};
    \node[kasten, below=.5cm of plant] (encoder) {Encoder: $f_k(x(0\!:\!\tau_k-1))$};
    \node[kasten, below=.8cm of encoder, align=center] (channel) {Uncertain\\ wiretap channel\\ $(\mathbf T_B^{n},\mathbf T_C^{n})$};
    \node[left=.7cm of channel] (aux1) {};
    \node[right=.7cm of channel] (aux2) {};
    \node[kasten, below=1.2cm of aux1] (estimator) {Estimator: $\varphi_k(b(0\!:\!t_k-1))$};
    \node[kasten, below=1.2cm of aux2] (eavesdropper) {Eavesdropper};
    \node[below=.5cm of estimator] (estimate) {$\hat x(\tau_{k-1}\!:\!\tau_k-1)$};
    \node[below=.4 of eavesdropper, align = center] (question) {large estimation error\\in d- or v-secure sense};
    
    \draw[pfeil] (noise) -> (plant);
    \draw[pfeil] (plant) -> (encoder);
    \draw[pfeil] (encoder) -> (channel) node[left, midway] {$a(t_{k-1}\!:\!t_k-1)$};
    \draw[pfeil] (channel) -| (estimator) node[left, near end] {$b(t_{k-1}\!:\!t_k-1)$};
    \draw[pfeil] (channel) -| (eavesdropper) node[right, near end] {$c(t_{k-1}\!:\!t_k-1$)};
     \draw[pfeil] (estimator) -> (estimate);
     \draw[pfeil] (eavesdropper) -> (question);
 \end{tikzpicture}
	\caption{The $k$-th step of a transmission scheme $(f_k,\varphi_k)_{k=0}^\infty$ with blocklength $n$.}\label{fig:setting}
\end{figure}
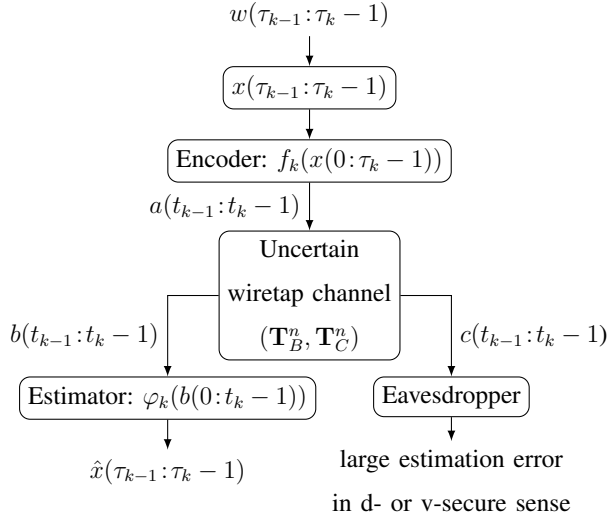

The concept is illustrated in Fig. \ref{fig:setting}. The encoding function $f_k$ takes the system path $x(0\!:\!\tau_k-1)$ until time $\tau_k-1$ as input and maps this into a codeword of length $n$. The blocks of new observations also have length $n$, except for the first one of length 1. Thus the initial state gets a special treatment, but this is a technical detail the reason of which will become clear in the proof of Theorem \ref{thm:main} below. We allow $f_k$ to be an uncertain channel for two reasons. One is that we do not have to distinguish between open and closed quantizing sets---if a path or state is on the boundary, we make an uncertain decision. The more important reason is that uncertain encoding has to be allowed in order for uncertain wiretap channels to achieve capacity, see Example \ref{ex:singletonnec} in Appendix \ref{app:channels}.

The decoder $\varphi_k$ takes the first $t_k$ outputs of $\mbf T_B$ and calculates an estimate of the states $x(\tau_{k-1}),\ldots,x(\tau_k-1)$ (where we set $\tau_{-1}=0$), which have not been estimated before.  When we define the performance criterion for a transmission scheme, it will be seen that by not allowing $\varphi_k$ to be an uncertain channel we do not lose generality.

Next we come to the definition of reliability and security of a transmission scheme $(f_k,\varphi_k)_{k=0}^\infty$. Every such transmission scheme induces the (uncertain) channels
	\[
		f_{0:k}:=f_0\times\cdots\times f_k,\qquad\varphi_{0:k}:=\varphi_0\times\cdots\times\varphi_k.
	\] 
Observe that, given a sequence $\hat x(0\!:\!\tau_k-1)$ of system estimates, i.e., of outputs of $\varphi_{0:k}$, we can write the set of system states which can generate this output sequence as $(f_{0:k}^{-1}\circ\mbf T_B^{-t_k}\circ\varphi_{0:k}^{-1})(\hat x(0\!:\!\tau_k-1))$.

Let $T$ be a positive integer or $\infty$. The $\infty$-norm of a real sequence $y(0\!:\!T)$ is given by
\begin{align*}
	&\lVert y(0\!:\!T)\rVert_\infty
	:=\begin{cases}
	                                   	\max_{0\leq t\leq T}\lvert y(t)\rvert&\text{if }T<\infty,\\
	                                   	\sup_{0\leq t<\infty}\lvert y(t)\rvert&\text{if }T=\infty.
	                                   \end{cases}
\end{align*}
For a set $\mc E\subset\mbb R^{T+1}$, where $T$ is a positive integer or infinity, we define its diameter by
\begin{align*}
	&\diam_{T+1}(\mc E)\\
	&:=\sup\{\lVert y(0\!:\!T)-y'(0\!:\!T)\rVert_\infty:y(0\!:\!T),y'(0\!:\!T)\in\mc E\}.
\end{align*}

\begin{defn}\label{def:reliability}
	The transmission scheme $(f_k,\varphi_k)_{k=0}^\infty$ is called \textit{reliable} if the estimation error is bounded uniformly in the estimates, i.e., if 
		there exists a constant $\kappa>0$ such that for every possible\footnote{Due to the application of the $\infty$-norm, the reliability criterion is a ``pointwise" criterion. Using $p$-norms of the form $\lVert y(0\!:\!T)\rVert_p:=(\sum_{t=0}^T\lvert y(t)\rvert^p)^{1/p}$ for some $1\leq p<\infty$ would always lead to an infinite estimation error if $\Omega>0$ and $\mbf T_B$ can transmit at most a finite number of messages in finite time, since the sequence $\lvert x(t)-\hat x(t)\rvert:t\geq 0$ would not tend to zero for all state sequences $x(0\!:\!\infty)$.} $\hat x(0\!:\!\infty)\in\ran(\varphi_{0:\infty}\circ\mbf T_B^\infty\circ f_{0:\infty})$, 
	\begin{equation}\label{eq:reldef}
		\sup_{k}\diam_{\tau_k}\bigl((f_{0:k}^{-1}\circ\mbf T_B^{-t_k}\circ\varphi_{0:k}^{-1})(\hat x(0\!:\!\tau_k-1))\bigr)\leq \kappa.
	\end{equation}
\end{defn}

\begin{rem}
	One would not gain anything by allowing the decoding functions $\varphi_k$ to be uncertain channels since this generalization could only increase the left-hand side of \eqref{eq:reldef}.
\end{rem}

A transmission scheme only defines a decoder at the output of the estimator's channel $\mbf T_B$. But every system path $x(0\!:\!\infty)$ also generates a sequence $c(0\!:\!\infty)\in \mbf T_C^\infty(f_{0:\infty}(x(0\!:\!\infty)))$ of outputs obtained by the eavesdropper. The two security criteria we define next require state information to be secure no matter how the eavesdropper further processes its channel output sequence. The first criterion just ensures that the eavesdropper's estimation error grows unbounded with time.

\begin{defn}\label{def:d-sec}
	The transmission scheme $(f_k,\varphi_k)_{k=0}^\infty$ is called \textit{d-secure} if there exists a function $\delta(k)$ with 
	\[
	 \diam_{\tau_k}\bigl((f_{0:k}^{-1}\circ\mbf T_C^{-t_k})(c(0\!:\!t_k-1))\bigr)\geq\delta(k)
	\]
	for all $c(0\!:\!\infty)\in\ran(\mbf T_C^{\infty}\circ f_{0:\infty})$ and $\delta(k)\rightarrow\infty$ as $k\rightarrow\infty$.
\end{defn}

Upon receiving any sequence $c(0\!:\!\infty)$ of channel outputs generated by a d-secure transmission scheme, the eavesdropper's estimate of the system path $x(0\!:\!\infty)$ that generated $c(0\!:\!\infty)$ grows to infinity\footnote{Note that d-security as defined via the $\infty$-norm is stronger than the analogous criteria with the $p$-norm instead of the $\infty$-norm for all $1\leq p<\infty$ because $\lVert x(0\!:\!\infty)\rVert_\infty\leq\lVert x(0\!:\!\infty)\rVert_p$.} uniformly in $c(0:\infty)$. Note that since $\mc X_{0:t}$ is bounded for every $t\geq 0$, the diameter of $(f_{0:k}^{-1}\circ\mbf T_C^{-t_k})(c(0\!:\!t_k-1))$ cannot be infinite for any $k$. Thus the eavesdropper's estimation error will always be finite, though increasingly large, in finite time.

Next one can ask the question how many system paths could be the possible generators of an  eavesdropper sequence $c(0\!:\!\infty)$. This is considered in the following secrecy criterion. For a set $\mc E$ of real sequences of finite length $T+1$ and $0\leq t\leq T$, we write $\mc E\vert_t:=\{x\in\mbb R:x=x(t)\text{ for some }x(0\!:\!T)\in\mc E\}$. The volume $\vol(\mc E')$ of a subset $\mc E'$ of the real numbers is measured in terms of the Lebesgue measure. 

\begin{defn}\label{def:v-sec}
  A transmission scheme $(f_k,\varphi_k)_{k=0}^\infty$ is called \textit{v-secure} if there exists a function $\nu(k)$ such that
  \[
    \vol((f_{0:k}^{-1}\circ\mbf T_C^{-t_k})(c(0\!:\!t_{k}-1))\rvert_{\tau_k-1})\geq\nu(k)
  \]
  for all $c(0\!:\!\infty)\in\ran(\mbf T_C^{\infty}\circ f_{0:\infty})$ and $\nu(k)\rightarrow\infty$ as $k\rightarrow\infty$.
\end{defn}

Like in the definition of d-security, we require uniform divergence to infinity. Since $\mc X_{0:t}$ is bounded for all $t\geq 0$, the volume in Definition \ref{def:v-sec} cannot be infinite in finite time.

\begin{rem}
Clearly, v-security implies d-security. The volume is measured at time $\tau_k-1$ because it would trivially tend to infinity if the $\tau_k$-dimensional volume of the set $(f_{0:k}^{-1}\circ\mbf T_C^{-t_k})(c(0\!:\!t_{k}-1))$ were measured. If the volume of the set of states tends to infinity along $\tau_k-1$ as $k\rightarrow\infty$, then the same holds for the volume measured at all other infinite, increasing sequences of time instances.
\end{rem}

\subsection{Results for Secure Estimation}

We first state a sufficient condition the uncertain wiretap channel has to satisfy in order for reliability as well as d- or v-security to be possible. 

\begin{theorem}\label{thm:main}
  There exists a transmission scheme which is reliable, d-secure and v-secure if $C_0(\mbf T_B,\mbf T_C)>\log\lambda$.
\end{theorem}

The proof of Theorem \ref{thm:main} can be found in Section \ref{sect:main-proof}. The transmission schemes applied there separate quantization/estimation from coding for uncertain wiretap channels by concatenating a quantizer defined below with a wiretap zero-error code. Note that the condition $C_0(\mbf T_B,\mbf T_C)>0$ is weak: Nair \cite{Nnst} proved that $C_0(\mbf T_B)>\log\lambda$ is sufficient and $C_0(\mbf T_B)\geq\log\lambda$ is necessary to achieve reliability. Thus by Theorem \ref{thm:zescap}, the additional requirement in Theorem \ref{thm:main} is nothing but $C_0(\mbf T_B,\mbf T_C)>0$. This is the minimal condition one would expect to be necessary to also achieve security. For general $(\mbf T_B,\mbf T_C)$ we do not know that $C_0(\mbf T_B,\mbf T_C)>0$ really has to be satisfied for secure estimation to be possible.

For injective channels, however, the condition from Theorem \ref{thm:main} is ``almost'' necessary to achieve reliability and d-security, hence also for v-security.

\begin{theorem}\label{thm:estconv}
 If $\mbf T_B$ is injective and $\mc C$ finite, then the existence of a reliable and d-secure transmission scheme implies $\card\mc A\geq\lambda$ and $C_0(\mbf T_B,\mbf T_C)>0$. 
\end{theorem}

The proof of this theorem can be found in Section \ref{subsect:conv}. Since $\mbf T_B$ is injective, the condition $\card\mc A\geq\lambda$ means nothing but $C_0(\mbf T_B)\geq\log\lambda$. As noted above, $C_0(\mbf T_B)\geq\log\lambda$ was shown by Nair \cite{Nnst} to follow from reliability for general uncertain channels. The additional condition $C_0(\mbf T_B,\mbf T_C)>0$, which follows from d-security, implies $C_0(\mbf T_B,\mbf T_C)\geq\log\lambda$ by Theorem \ref{thm:zescap}. The problem of finding a tight necessary condition for secure estimation over general uncertain wiretap channels $(\mbf T_B,\mbf T_C)$ remains open. We conjecture that it depends on a criterion for $C_0(\mbf T_B,\mbf T_C)$ to be positive. We only have such a criterion in the case that $\mbf T_B$ is injective from Theorem \ref{thm:injch}.

As a refinement of Theorem \ref{thm:main}, we have a closer look at the exponential rate at which the estimation error or the volume of the set of states at a given time which are possible according to the eavesdropper's information tend to infinity. The higher the speed of divergence, the higher is the degree of security. 

\begin{lem}\label{lem:divcoeff}
	There exists a reliable transmission scheme $(f_k,\varphi_k)_{k=0}^\infty$ such that for every $c(0\!:\!\infty)\in\ran(\mbf T_C^\infty\circ f_{0:\infty})$ there exist system paths $x(0\!:\!\infty),x'(0\!:\!\infty)\in(f_{0:\infty}^{-1}\circ\mbf T_C^{-\infty})(c(0\!:\!\infty))$ satisfying
	\begin{equation}\label{eq:divcoeff-cond}
		\lim_{t\rightarrow\infty}\frac{\log\lVert x(0\!:\!t)-x'(0\!:\!t)\rVert_\infty}{t}=\log\lambda.
	\end{equation}
\end{lem}

This lemma is proved in Section \ref{sect:divcoeff-proof}. Clearly, $\log\lambda$ is the largest exponential rate at which two trajectories can diverge. For v-security, the speed of increase of the volume of the set of possible states according to the eavesdropper's information will in general increase at an exponential rate smaller than $\log\lambda$. 

\begin{lem}\label{lem:volrate}
  For every zero-error wiretap $(n,M,\gamma)$-code $\mbf F$, upon setting
  \begin{equation}\label{eq:raten}
   \frac{\log M}{n}=:R,\quad\frac{\log\gamma}{n}=:\Gamma,
  \end{equation}
there exists a reliable transmission scheme $(f_k,\varphi_k)_{k=0}^\infty$ with blocklength $n$ such that for all $c(0\!:\!\infty)\in\ran(\mbf T_C^{\infty}\circ f_{0:\infty})$, 
  \begin{align}
    &\lim_{k\rightarrow\infty}\frac{\log\vol((f_{0:k}^{-1}\circ\mbf T_C^{-t_k})(c(0\!:\!t_k-1))\vert_{\tau_k-1})}{\tau_k}\notag\\
    &\quad\geq 
    \begin{cases}\label{eq:vol-div-raten}
      \Gamma+\log\lambda-R &\text{if }\Omega=0,\\
      \frac{\Gamma\log\lambda}{R+2\log\lambda+\varepsilon_n} &\text{if }\Omega>0,
    \end{cases}
  \end{align}
  where $\varepsilon_n=\varepsilon_n(R,\lambda)$ is positive and $\varepsilon_n\rightarrow0$ as $n\rightarrow\infty$. 
\end{lem}

This lemma is proved in Section \ref{sect:main-proof}. For $\Omega=0$, a positive rate is achievable by choosing $R<\Gamma+\log\lambda$. Lemmas \ref{lem:divcoeff} and \ref{lem:volrate} are discussed in detail in Section \ref{sect:numex}.

\section{Quantizer Analysis}\label{sect:quantana}

Both Lemmas \ref{lem:divcoeff} and \ref{lem:volrate} follow from analyzing the transmission scheme we apply in the proof of Theorem \ref{thm:main}. For proving Theorem \ref{thm:main}, we separate quantization/estimation from channel coding. Next, we will therefore describe the quantizer used in the proof of Theorem \ref{thm:main}. More precisely, we analyze the behavior of the system \eqref{eq:dynsyst} with an appropriate quantization of every single state $x(t)$. Later, when concatenating the quantizer with a channel code of blocklength $n>1$, we will use an analogous quantizer for the $n$-sampled version of \eqref{eq:dynsyst}.

\begin{defn}\label{def:quant}
Consider the system \eqref{eq:dynsyst} and let $M\geq 2$ be an integer, called the \emph{number of quantizer levels}. Let $\hat x(m(0\!:\!-1))$ be the mid point of $\mc I(m(0\!:\!-1)):=\mc I_0$. For every integer $t\geq 0$ and every sequence $m(0\!:\!t)\in\{0,\ldots,M-1\}^{t+1}$, we then recursively set
\begin{align}
	&\mc P(m(0\!:\!t)):=\mc I(m(0\!:\!t-1))_{\min}\notag\\
	&\qquad\qquad+\frac{\lvert\mc I(m(0\!:\!t-1))\rvert}{M}\left[m(t),m(t)+1\right],\label{eq:Pmn}\\
	&\hat x(m(0\!:\!t)):=\text{mid point of }\mc P(m(0\!:\!t)),\label{eq:hatx}\\
	&\mc I(m(0\!:\!t)):=\lambda\mc P(m(0\!:\!t))+\left[-\frac{\Omega}{2},\frac{\Omega}{2}\right]\label{eq:A(n),B(n)}.
\end{align}
(in \eqref{eq:Pmn}, recall our notation for intervals). Finally we define for every $t\geq 0$ the \textit{$t$-th quantizer channel}, an uncertain channel $\mbf Q_{t}$ which maps any message sequence $m(0\!:\!t-1)$ and any $x(t)\in\mc I(m(0\!:\!t-1))$ to an element of
\begin{equation}\label{eq:quantt}
  \mbf Q_{t}(x(t),m(0\!:\!t\!-\!1))\!=\!\{m\!:\!x(t)\!\in\!\mc P(m(0\!:\!t\!-\!1),m)\}.
\end{equation}
The sets $\mc P(\cdot)$ will be referred to as \textit{quantizer intervals}. The numbers $0,\ldots,M-1$ are \textit{messages}. Equations \eqref{eq:Pmn}-\eqref{eq:quantt} define the \textit{quantizer of the system \eqref{eq:dynsyst} with $M$ quantizer levels}.
\end{defn}

Every state sequence $x(0\!:\!\infty)$ generates a message sequence $m(0\!:\!\infty)$ via the uncertain channels $\mbf Q_{t}$. Assume that the state sequence $x(0\!:\!t-1)$ has generated message sequence $m(0\!:\!t-1)$ until time $t-1$. The interval $\mc I(m(0\!:\!t-1))$ consists of all states $x(t)$ which are possible in the next time step. Upon observation of $x(t)$, the message $m(t)$ is generated as an element\footnote{$m(t)$ is not determined deterministically from $x(t)$ and $m(0\!:\!t-1)$ because in this way we can have all intervals $\mc P(m(0\!:\!t))$ closed. Note that $\card\mbf Q_{t}(x(t),m(0\!:\!t-1))\geq 2$ only if $x(t)$ is on the boundary of two neighboring quantizer intervals.} of $\mbf Q_{t}(x(t),m(0\!:\!t-1))$. From the sequence $m(0\!:\!t)$ one can then infer that $x(t)\in\mc P(m(0\!:\!t))$. Accordingly, the estimate of $x(t)$ is $\hat x(m(0\!:\!t))$. Note that for every message sequence $m(0\!:\!\infty)$ there exists a system path $x(0\!:\!\infty)$ which generates $m(0\!:\!\infty)$.

Most of the quantizer analysis we do in the following serves the proof of Lemma \ref{lem:volrate}. We are interested in the disjointness of quantizer intervals at a given time in order to find a lower bound on the volume of the set of states which are possible according to the eavesdropper's information: If a set of quantizer intervals at a common time instant is disjoint, the volume covered by their union equals the sum over their individual volumes. Thus two questions need to be answered: 1) What is the volume of a quantizer interval? 2) How many disjoint quantizer intervals are there (from the eavesdropper's view)? An answer to the first question is the following lemma.

\begin{lem}\label{thm:estdyn}
	If $\lambda\neq M$, then for every $t\in\mbb N$ and $m(0\!:\!t)\in\{0,\ldots,M-1\}^{t+1}$ we have
	\begin{align}\label{eq:intlength}
		\lvert\mc P(m(0\!:\!t))\rvert=\frac{\lambda^t}{M^t}\left(\frac{\lvert\mc I_0\rvert}{M}-\frac{\Omega}{M-\lambda}\right)+\frac{\Omega}{M-\lambda}.
	\end{align}
	In particular, we have $\sup_t\lvert\mc P(m(0\!:\!t))\rvert<\infty$ for every infinite message sequence $m(0\!:\!\infty)$ if $\lambda<M$. In that case
	\[
		\sup_{t\geq 0}\lvert\mc P(m(0\!:\!t))\rvert=\max\left\{\frac{\lvert\mc I_0\rvert}{M},\frac{\Omega}{M-\lambda}\right\}.
	\]
	Further, the length of $\mc P(m(0\!:\!t))$ only depends on $t$, not on $m(0\!:\!t)$. Thus we can define
	\begin{equation}\label{eq:ell-def}
		\ell_{t}:=\lvert\mc P(m(0\!:\!t))\rvert.
	\end{equation}
\end{lem}

The proof can be found in Appendix \ref{app:quantana}. Lemma \ref{thm:estdyn} not only is useful in the security analysis, but it also essentially establishes reliability for $M>\lambda$, a result which of course is not surprising in view of the existing literature. Concerning question 2), life is simple in the case $\Omega=0$ because of the following lemma.

\begin{lem}\label{lem:nullklar}
	If $\Omega=0$, then at each time $t\geq 0$, the interiors of the intervals $\mc P(m(0\!:\!t))$ are disjoint, where $m(0\!:\!t)$ ranges over $\{0,\ldots,M-1\}^{t+1}$.
\end{lem}

For the proof, see Appendix \ref{app:quantana}. Thus at time $t$, we have $M^{t+1}$ disjoint quantizer intervals of the same length. If $\Omega>0$, then the situation is more complicated: Quantizer intervals belonging to different message sequences of the same length can overlap. This is the reason for the two different lower bounds on the rate of volume increase in \eqref{eq:vol-div-raten}.

\begin{example}\label{ex:ueberlappung}
	Consider the system \eqref{eq:dynsyst} with $\lambda=1.2$, $\Omega=.1$, $\mc I_0=[-1,1]$ and its quantizer with $M=3$. Then $\mc P(0)=[-1,-1/3]$ and $\mc P(1)=[-1/3, +1/3]$. In the next time step, one has 
	\[
		\mc P(0, 1)=\left[-.6,-.35\right],\quad\mc P(1, 0)=\left[-.45, -.15\right],
	\]
	so $\mc P(0,1)$ and $\mc P(1,0)$ are not disjoint. The closer a state $x(t)$ is to the origin (and the larger $t$), the more paths there are which can be in this particular state at time $t$.
\end{example}

Example \ref{ex:ueberlappung} shows that one can only hope to obtain disjoint quantizer sets for a strict subset of all message sequences. To find such a subset, we derive an important formula for the sequence $\hat x(m(0\!:\!\infty))$ given a message sequence $m(0\!:\!\infty)$.

\begin{lem}\label{thm:mpdyn}
	Consider the system \eqref{eq:dynsyst} and consider the quantizer for \eqref{eq:dynsyst} with $M$ quantizer levels. Let $m(0\!:\!\infty)$ be a message sequence. Then for every $t=0,1,2,\ldots$
	\begin{align}\label{eq:estform}
		&\!\!\hat x(m(0\!:\!t))\notag\\
		&\!\!=\lambda^t
		\biggl\{
		\hat x(m(0\!:\!-1))\notag\\
		&\!\!+\!\frac{1}{2}\!\sum_{i=0}^{t}\!\left(\!\frac{\Omega M}{M\!-\!\lambda}\!\left(\frac{1}{\lambda^i}\!-\!\frac{1}{M^i}\!\right)\!+\!\frac{\lvert\mc I_0\rvert}{M^i}\!\right)\!\!\left(\frac{2m(i)\!+\!1}{M}\!-\!1\right)\!\biggr\}.
	\end{align}
\end{lem}

See Appendix \ref{app:quantana} for the proof. In order to find disjoint quantizer intervals, the idea is to look at the distance between points $\hat x(m(0:t))$ and $\hat x(m'(0:t))$ and ask how the distances between the estimate sequences will evolve in the future.  

\begin{lem}\label{thm:estandsecwithout}
	Assume that $M>\lambda$. Let $m(0\!:\!\infty),m'(0\!:\!\infty)$ be two message sequences and let $T\geq 0$. If 
	\begin{equation}\label{eq:voraussohne}
		\lvert\hat x(m(0\!:\!T))-\hat x(m'(0\!:\!T))\rvert\geq\frac{\Omega}{M-\lambda}\frac{M-1}{\lambda-1}+\ell_T,
	\end{equation}
	then for every $t\geq 0$, the interiors of the intervals $\mc P(m(0\!:\!T+t))$ and $\mc P(m'(0\!:\!T+t))$ are disjoint.
\end{lem}

The proof can be found in Appendix \ref{app:quantana}. Finally, assume that at each time instant at least $\gamma$ different messages are possible according to the eavesdropper's view. For every $t\geq 0$ let $\mc M_t:=\{m_{t,1}<m_{t,2}<\ldots<m_{t,\gamma}\}\subseteq\{0,\ldots,M-1\}$ be a subset of the possible messages at time $t$ which has exactly $\gamma$ elements. In particular, $\mc M_t$ may differ from $\mc M_{t'}$ for $t\neq t'$. Now fix a $T\geq 1$. For $j\geq 1$ and $\xi(1\!:\!j)\in\{1,\ldots,\gamma\}^j$, we define the message sequence $m_{\xi(1:j)}(0\!:\!jT-1)$ by
	\[
		m_{\xi(1:j)}(s)=m_{s,\xi(i)}\in\mc M_s
	\]
if $1\leq i\leq j$ and $(i-1)T\leq s\leq iT-1$. On the $j$-th block of times $(j-1)T,\ldots,jT-1$, the sequences $m_{\xi(1:j)}(0\!:\!jT-1)$, where $\xi(1\!:\!j-1)$ is kept fixed and $\xi(j)$ ranges over $\{1,\ldots,\gamma\}$, are an ordered set of $\gamma$ message sequences with the order induced by componentwise ordering. The corresponding quantizer intervals $\mc P(m_{\xi(1:j)}(0\!:\!jT-1))$, where $1\leq \xi(j)\leq\gamma$, will therefore diverge due to the instability of the system \eqref{eq:dynsyst}. The following lemma is proved in Appendix \ref{app:quantana}.

\begin{lem}\label{lem:vol}
	Let $\Omega>0$ and $M>\lambda$ and choose a $T\in\mbb N$ satisfying
	\begin{equation}\label{eq:T}
		T\geq 1+\frac{\log(M-1)+\log(M+\lambda-1)-\log(M-\lambda)}{\log\lambda}
	\end{equation} 
	Then for every $j\geq 1$, the interiors of the sets $\mc P(m_{\xi(1:j)}(0\!:\!jT-1))$, where $\xi(1\!:\!j)$ ranges over $\{1,\ldots\gamma\}^j$, are disjoint.
\end{lem}

Thus we have obtained a lower bound on the number of disjoint quantizer intervals at times $t=jT-1$, for positive $j$. This will be sufficient when we put everything together in the next section to prove v-security and obtain the lower bound of Lemma \ref{lem:volrate} for the case $\Omega>0$.

\section{Secure Estimation -- Proofs}\label{sect:secestres}

\subsection{Proof of Theorem \ref{thm:main} and Lemmas \ref{lem:divcoeff} and \ref{lem:volrate}}\label{sect:main-proof}

\subsubsection{Definition of the Transmission Scheme}

We start by defining a transmission scheme $(f_k,\varphi_k)_{k=0}^\infty$. We choose its blocklength $n$ such that $M:=N_{(\mbf T_B,\mbf T_C)}(n)>\lambda^n$, which is possible because $C_0(\mbf T_B,\mbf T_C)>\log\lambda$. Let $\gamma\geq 2$ be chosen such that there exists a zero-error wiretap $(n,M,\gamma)$-code $\mbf F$. 

Since we use the channel in blocks of length $n$, we also observe the system only at intervals of length $n$. If we look at the outputs of \eqref{eq:dynsyst} at times $0,n,2n,\ldots$, we obtain a new dynamical system which satisfies
\begin{subequations}\label{eq:n-smpl-normal}
\begin{align}
	x\e{n}(k+1)&=\lambda^nx^{(n)}(k)+w^{(n)}(k),\\
	x\e{n}(0)&\in\mc I_0,
\end{align}
\end{subequations}
where 
\[
	w^{(n)}(k)=\sum_{j=0}^{n-1}\lambda^{n-j-1}w(kn+j).
\]
Note that $w\e{n}(k)$ is a nonstochastic disturbance in the range $[-\Omega^{(n)}/2,\Omega^{(n)}/2]$ for
\begin{equation}\label{eq:n-Omega-normal}
	\Omega\e{n}=\frac{\Omega}{\lambda-1}(\lambda^{n}-1).
\end{equation}
Therefore the quantizer for \eqref{eq:n-smpl-normal} with $M$ quantization levels is well-defined as in Definition \ref{def:quant} and all results derived in the previous section for \eqref{eq:dynsyst} and its quantizer carry over to \eqref{eq:n-smpl-normal} with the obvious modifications of the parameters.

We define the encoding and decoding functions of our transmission scheme by separating quantization/estimation from channel coding like it has been done frequently in settings without security, e.g., \cite{WB1}. For every $k\geq 0$, let $\mbf Q_k\e n$ be the $k$-th quantizer channel of the quantizer of \eqref{eq:n-smpl-normal} (see \eqref{eq:quantt}). The transmission scheme is defined by recursively concatenating the $\mbf Q_k\e n$ with $\mbf F$. We set $f_0(x(0))=\mbf F(\mbf Q_{0}\e n(x(0)))$ and for $k\geq 1$, assuming that the quantizer channels have produced the message sequence $m(0\!:\!k-1)$ so far, we set
\[
	f_k(x(0\!:\!\tau_k-1))=\mbf F(\mbf Q_{k}\e n(x\e n(k),m(0\!:\!k-1))).
\]
For the definition of the decoding functions, recall that $\mbf F$ is a zero-error code. Thus for every $k\geq 0$ and $b(0\!:\!t_k-1)\in\ran(\mbf T_B^{t_k}\circ\mbf F^{k+1})$, the set $(\mbf F^{-(k+1)}\circ\mbf T_B^{-t_k})(b(0\!:\!t_k-1))$ contains precisely one element, namely the message sequence $m(0\!:\!k)$ sent by the encoder. The $0$-th decoding function has a 1-dimensional output which is defined by $\varphi_0(b(0\!:\!t_0-1))=\hat x\e n((\mbf F^{-1}\circ\mbf T_B^{-t_0})(b(0\!:\!t_0-1)))$. Here $\hat x\e n(m(0:k))$ for any $m(0:k)$ is the mid point of the quantizer interval $\mc P\e n(m(0:k))$ belonging to the quantizer of \eqref{eq:n-smpl-normal}. For $k\geq 1$, the output of the $k$-th decoding function $\varphi_k$ is $n$-dimensional. If, with a little abuse of notation, we write $\varphi_k(b(0\!:\!t_k-1))=:(\hat x_{\tau_{k-1}}(b(0\!:\!t_k-1)),\ldots,\hat x_{\tau_k-1}(b(0\!:\!t_k-1)))$, then we set 
\[
	\hat x_{\tau_k-1}(b(0\!:\!t_k-1))=\hat x\e n((\mbf F^{-(k+1)}\circ\mbf T_B^{-t_k})(b(0\!:\!t_k-1))).
\]
Since \eqref{eq:dynsyst} does not grow to infinity in finite time, the values $\hat x_{\tau_{k-1}}(b(0\!:\!t_k-1)),\ldots,\hat x_{\tau_k-2}(b(0\!:\!t_k-1))$ can be defined in an arbitrary way as long as their distance from $\hat x_{\tau_k-1}(b(0\!:\!t_k-1))$ is uniformly bounded in $k$ and $b(0\!:\!\infty)$. 

\subsubsection{Reliability}

Although it is not surprising and well-known in the literature, for completeness we show the reliability of the transmission scheme. Since the states of \eqref{eq:dynsyst} cannot diverge to infinity in finite time, we only need to make sure that the estimation errors at the observation times $\tau_0-1,\tau_1-1,\ldots$ are bounded. To see this, let $k\geq 0$ and $m(0\!:\!k)$ any message sequence and observe that
\[
  (f_{0:k}^{-1}\circ\mbf T_B^{-t_k}\circ\varphi_{0:k}^{-1})(\hat x\e n(m(0\!:\!k)))\vert_{\tau_k-1}=\mc P\e n(m(0\!:\!k)).
\]
Since $M>\lambda^n$, the length of $\mc P\e n(m(0\!:\!k))$ is bounded by Lemma \ref{thm:estdyn}. This shows that the transmission scheme is reliable.

\subsubsection{d-Security and Lemma \ref{lem:divcoeff}}\label{sect:divcoeff-proof}

Let $c(0\!:\!\infty)\in\ran(\mbf T_C^\infty\circ f_{0:\infty})$. Let $m(0)\neq m'(0)\in\mbf F^{-1}(\mbf T_C^{-n}(c(0\!:\!t_0-1)))$ and $m(k)\in\mbf F(\mbf T_C^{-n}(c(t_{k-1}\!:\!t_k-1)))$. Then there are two system trajectories $x(0\!:\!\infty),x'(0\!:\!\infty)$ such that $x(\tau_k-1)=\hat x\e n(m(0\!:\!k))$ and $x'(\tau_k-1)=\hat x\e n(m'(0)m(1\!:\!k))$ for all $k\geq 0$. With Lemma \ref{thm:mpdyn} one immediately sees that $x(0\!:\!\infty)$ and $x'(0\!:\!\infty)$ diverge at exponential rate $\log\lambda$. Thus $x(0\!:\!\infty),x'(0\!:\!\infty)$ satisfy \eqref{eq:divcoeff-cond}. This proves Lemma \ref{lem:divcoeff} and the achievability of d-security.

\subsubsection{v-Security and Lemma \ref{lem:volrate}}\label{sect:volrate-proof}

For the proof of v-security of the transmission scheme, we consider the two subcases $\Omega=0$ and $\Omega>0$. We first assume $\Omega=0$, hence $\lvert\mc I_0\rvert>0$. In this case, hardly anything remains to be proved. By Lemma \ref{lem:nullklar}, for given $k\geq 0$, the interiors of all $\mc P\e n(m(0\!:\!k))$ are disjoint. Now assume that the eavesdropper receives the sequence $c(0\!:\!t_k-1)$. Since $\mbf F$ is a $(n,M,\gamma)$-code, $	\card(\mbf F^{-(k+1)}\circ\mbf T_C^{-t_k})(c(0\!:\!t_k-1))\geq\gamma^{k+1}$. Hence 
\begin{align*}
	&\vol((f_{0:k}^{-1}\circ\mbf T_C^{-t_k})(c(0\!:\!t_k-1))\vert_{\tau_k-1})\\
	&\quad=\sum_{m(0:k)\in(\mbf F^{-(k+1)}\circ\mbf T_C^{-t_k})(c(0:t_k-1))}\ell_k\e n\\
	&\quad\quad\geq\gamma^{k+1}\ell_k\e n=\left(\frac{\gamma\lambda^n}{M}\right)^k\frac{\gamma\lvert\mc I_0\rvert}{M},
\end{align*}
where $\ell_k\e n$ is the length of the quantizer intervals at time $k$ of the quantizer of \eqref{eq:n-smpl-normal}. This gives the possibly negative growth rate $(\log\gamma)/n+\log\lambda-(\log M)/n$, as claimed in Lemma \ref{lem:volrate} for the case $\Omega=0$. Since $(\log M)/n$ can be chosen  strictly smaller than $(\log\gamma)/n+\log\lambda$, this also proves that v-security is achievable for $\Omega=0$, and thus completes the proof of Theorem \ref{thm:main} for the case $\Omega=0$.

Next we assume that $\Omega>0$. Define 
\[
	T\e n:=\left\lceil 1+\frac{\log M}{n\log\lambda}+\frac{\log(M+\lambda^n)-\log(M-\lambda^n)}{n\log\lambda}\right\rceil.
\]
Choose a $j\geq 1$ and set $k(j):=jT\e n-1$. Let $c(0\!:\!t_{k(j)}-1)$ be an eavesdropper output sequence. Then by choice of $\mbf F$
\begin{equation}\label{eq:possin}
	\card(\mbf F^{-(k(j)+1)}\circ\mbf T_C^{-t_{k(j)}})(c(0\!:\!t_{k(j)}-1))\geq\gamma^{k(j)+1}.
\end{equation}
$T\e n$ satisfies \eqref{eq:T} for \eqref{eq:n-smpl-normal}. By Lemma \ref{lem:vol} applied to \eqref{eq:n-smpl-normal}, within the set on the left-hand side of \eqref{eq:possin}, the $\gamma^j$ message sequences of the form $m_{\xi(1:j)}(0\!:\!k(j))$ produce sets $\mc P\e n(m_{\xi(1:j)}(0\!:\!k(j)))$ with disjoint interiors. Therefore
\begin{align}
	&\vol((f_{0:k(j)}^{-1}\circ\mbf T_C^{-t_{k(j)}})(c(0\!:\!t_{k(j)}-1))\vert_{\tau_{k(j)}-1})\notag\\
	&\geq\sum_{\xi(1:j)\in\{1,\ldots,\gamma\}^j}\ell_{k(j)}\e n=\gamma^j\ell_{k(j)}\e n.\label{eq:rateroh}
\end{align}
Since $\ell_{k(j)}\e n$ tends to a constant as $j$ tends to infinity, the asymptotic rate of volume growth is lower bounded by
\[
  \lim_{k\rightarrow\infty}\frac{\log(\gamma^j\ell_{k(j)}\e n)}{\tau_k}=\frac{\log \gamma}{nT\e n}.
\]
With the notation \eqref{eq:raten} and setting 
\begin{gather*}
  \varepsilon_n:=\frac{\log(M+\lambda^n)-\log(M-\lambda^n)}{n},
\end{gather*}
we obtain
\begin{align*}
  \frac{\log\gamma}{nT\e n}\geq\frac{\Gamma\log\lambda}{R+2\log\lambda+\varepsilon_n}.
\end{align*}
Clearly, $\varepsilon_n$ is positive and tends to 0 as $n$ tends to infinity. This proves that v-security can be achieved in the case $\Omega>0$ as well, and at the rate claimed in Lemma \ref{lem:volrate}. Altogether, this completes the proof of Theorem \ref{thm:main} and Lemmas \ref{lem:divcoeff} and \ref{lem:volrate}.

\subsection{Proof of Theorem \ref{thm:estconv}}\label{subsect:conv}

Assume that $\mbf T_B$ is injective and $\mc C$ is finite. Let $(f_k,\varphi_k)_{k=0}^\infty$ be a reliable and d-secure transmission scheme with blocklength $n$. In particular, choose $\kappa>0$ in such a way that \eqref{eq:reldef} is satisfied for every possible sequence of estimates $\hat x(0\!:\!\infty)$. The necessity of $C_0(\mbf T_B)\geq\log\lambda$ was shown in \cite{Nnst}. Due to the injectivity of $\mbf T_B$, this condition can be reformulated as $\card\mc A\geq\lambda$. It remains to show that $C_0(\mbf T_B,\mbf T_C)>0$. 

By the uniform divergence requirement in the definition of d-security, it is possible to choose a $k$ such that
\begin{equation}\label{eq:unrel}
	\diam_{\tau_k}((f_{0:k}^{-1}\circ\mbf T_C^{-t_k})(c(0\!:\!t_k-1)))>\kappa
\end{equation}
for every $c(0\!:\!t_k-1)\in\ran(\mbf T_C^{t_k}\circ f_{0:k})$. Let $\tilde c(0\!:\!t_k-1)\in\ran(\mbf T_C^{t_k}\circ f_{0:k})$. Recursively, we define the sets $\mc T_0(\tilde c(0\!:\!t_k-1)):=\ran(f_{0:k})\cap\mbf T_C^{-t_k}(\tilde c(0\!:\!t_k-1))$ and $\mc T_j(\tilde c(0\!:\!t_k-1)):=\ran(f_{0:k})\cap(\mbf T_C^{-t_k}\circ\mbf T_C^{t_k})(\mc T_{j-1}(\tilde c(0\!:\!t_k-1)))$ for $j\geq 1$. Let $j_*$ be the maximal $j$ which satisfies\footnote{Without going into the details, we would like to mention here that $\mc T_{j_*}(c(0\!:\!t_k-1))$ is an equivalence class in the taxicab partition of the joint range of $f_{0:k}$ and the corresponding outputs of $\mbf T_C$, see \cite{Nnst}.\label{fn:taxicab}} $\mc T_j(\tilde c(0\!:\!t_k-1))\supsetneq\mc T_{j-1}(\tilde c(0\!:\!t_k-1))$.  If $a_0(0\!:\!t_k-1),\ldots,a_{M-1}(0\!:\!t_k-1)$ is an enumeration of the elements of $\mc T_{j_*}(\tilde c(0\!:\!t_k-1))$, then the $(M,t_k)$-code $\mbf G_k$ defined by $\mbf G_k(m)=\{a_m(0\!:\!t_k-1)\}$ is a zero-error code. This is due to the injectivity of $\mbf T_B$. 

But $\mbf G_k$ even is a wiretap zero-error code. To show this, let $c(0\!:\!t_k-1)\in\ran(\mbf T_C^{t_k}\circ\mbf G_k)$. The definition of $j_*$ implies that $\mbf T_C^{-t_k}(c(0\!:\!t_k-1))\subseteq\mc T_{j_*}(\tilde  c(0\!:\!t_k-1))=\ran(\mbf G_k)$. Due to \eqref{eq:unrel} and since $(f_k,\varphi_k)_{k=0}^\infty$ satisfies \eqref{eq:reldef}, we have $\card(\mbf G_k^{-1}\circ\mbf T_C^{-t_k})(c(0\!:\!t_k-1))=\card\mbf T_C^{-t_k}(c(0\!:\!t_k-1))\geq 2$. Hence $c(0\!:\!t_k-1)$ can be generated by at least two different messages. This implies that $\mbf G_k$ also is a wiretap zero-error code, hence $C_0(\mbf T_B,\mbf T_C)>0$.

\section{Discussion: d- and v-Security}\label{sect:numex}

We have a closer look at d- and v-security, in particular the rates derived in Lemmas \ref{lem:divcoeff} and \ref{lem:volrate}. First consider the system \eqref{eq:dynsyst} with $\Omega=0$. Let $(\mbf T_B,\mbf T_C)$ be any uncertain wiretap channel and $\mbf F$ an $(n,M,\gamma)$-code for $(\mbf T_B,\mbf T_C)$. Then the proof of Lemma \ref{lem:volrate} shows that the lower bound on the right-hand side of \eqref{eq:vol-div-raten} is tight. On the other hand, the growth rate $\log\lambda$ of the eavesdropper's estimation error derived in Lemma \ref{lem:divcoeff} will in general be strictly larger. This means that the set $(f_{0:k}^{-1}\circ\mbf T_C^{-t_k})(c(0:t_k-1))$ is not connected, i.e., it has holes.

\begin{figure}
 \includegraphics[width = \linewidth]{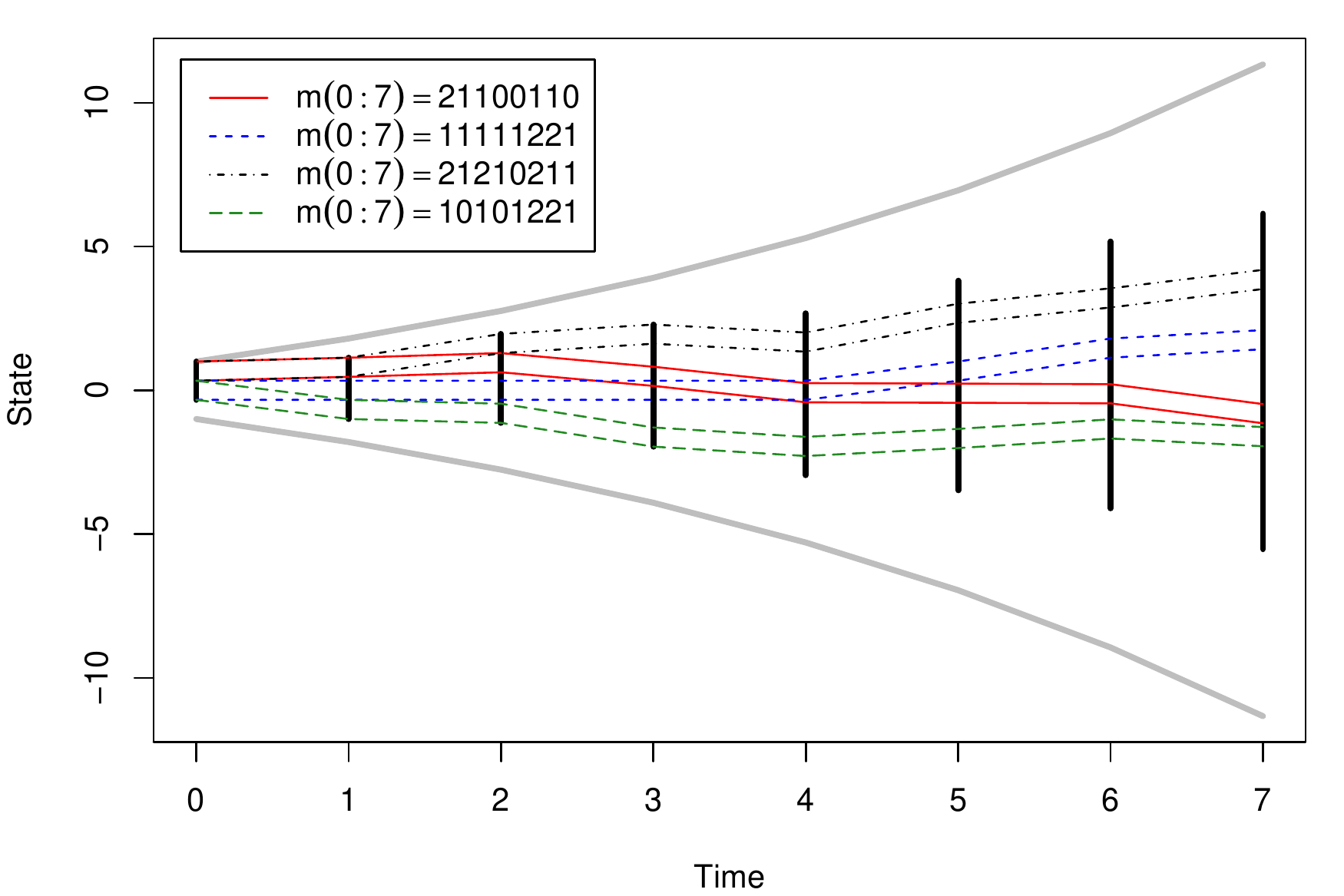}
 \caption{The state space of \eqref{eq:dynsyst} with parameters as in the text. The thick grey lines mark the outer bounds of the state space. For the received eavesdropper sequence $c(0\!:\!7)$ as in the text, the vertical black lines show the set of states which are possible according to the eavesdropper's view. Further, for four possible message sequences $m(0:7)$, the evolution of the corresponding $\mc P(m(0:7))$ is shown for illustration purposes.}\label{fig:illustration}
\end{figure}

If $\Omega>0$, we have seen in Example \ref{ex:ueberlappung} and the proof of Lemma \ref{lem:volrate} that the situation is more complicated than for $\Omega=0$. For an illustration, let $(\mbf T_B,\mbf T_C)$ and $\mbf F$ be the channel and code from Fig. \ref{fig:uncch-code}(b). Assume the system \eqref{eq:dynsyst} with $\lambda = 1.2, \mc I_0=[-1,1]$ and $\Omega =1.2$. As in the proof of Theorem \ref{thm:main}, we construct a blocklength-1 transmission scheme $(f_k,\varphi_k)_{k=0}^\infty$ by concatenating the quantizer for \eqref{eq:dynsyst} with $\mbf F$ by mapping the quantizer message $m$ to $\mbf F(m)$. For example, if $x(0)\in[1/3,1]$, the quantizer outputs message 2, which $\mbf F$ maps to the set $\mbf F(2)= \{a_4\}$. Sending $a_4$ through $\mbf T_C$ generates the output $c_2$, from which the eavesdropper concludes that message 1 or 2 has been sent. By choice of parameters, the length of the quantizer intervals remains constant over time. Fig. \ref{fig:illustration} illustrates this situation under the assumption that the eavesdropper receives the symbols $c(0\!:\!7)=c_2c_1c_2c_1c_1c_2c_2c_1$. There are $2^8$ possible message sequences from the eavesdropper's point of view, one of which corresponds to the actual sequence generated by the quantizer. Notice the growth of $\vol((f_{0:7}^{-1}\circ\mbf T_C^{-8})(c(0\!:\!7)))$, which also implies the growth of the eavesdropper's estimation error in the sense of d-security. Further observe how quantizer intervals overlap and even ``cross paths''.

Generally, if $\Omega>0$ and $\Gamma=R$, then the eavesdropper has no information about the transmitted message, and $\vol((f_{0:k}^{-1}\circ\mbf T_C^{-t_k})(c(0\!:\!t_k-1)))$ grows at rate $\log\lambda$. The ratio of the left- and the right-hand side of \eqref{eq:vol-div-raten} tends to 1 as $\lambda\searrow1$. Thus the lower bound of Lemma \ref{lem:volrate} is asymptotically tight for $\lambda$ tending to the boundary of the instability region.

\begin{figure}
  \includegraphics[width = \linewidth]{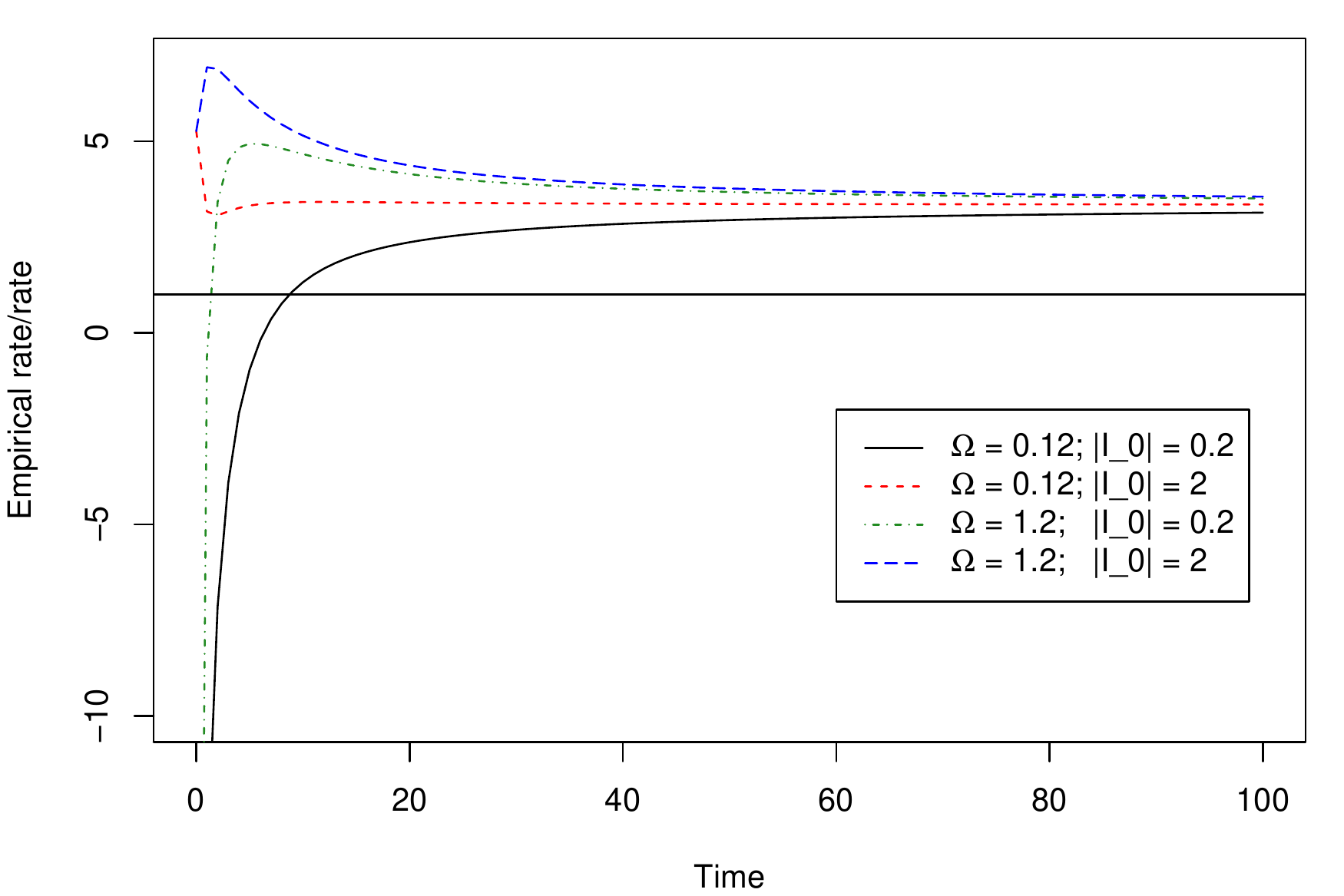}
  \caption{The ratio of the left- and right-hand side of \eqref{eq:vol-div-raten} for different combinations of $\Omega$ and $\mc I_0$, with other parameters as in the text.}\label{fig:Volumenrechnung}
 \end{figure}

Moreover, the lower bound \eqref{eq:vol-div-raten} for $\Omega>0$ is independent of $\Omega$ and of $\mc I_0$. This behavior can be expected by the asymptotic dominance of $\lambda$ in the system dynamics. Fig.~\ref{fig:Volumenrechnung} shows numerical evidence for the correctness of this independence. For the system parameters, we fix $\lambda=1.2$ and consider four variations of $\Omega$ and $\mc I_0$ as shown in Fig. \ref{fig:Volumenrechnung}. We assume the same uncertain wiretap channel as in Fig. \ref{fig:illustration} and apply the same  blocklength-1 transmission scheme. Because of the symmetry of the channel and the transmission scheme, $\vol((f_{0:k}^{-1}\circ\mbf T_C^{-(k+1)})(c(0\!:\!k)))$ is independent of the eavesdropper's received sequence and can be calculated in closed form. For each of the four combinations of $\Omega$ and $\mc I_0$ we plot the ratio of the left-hand side of \eqref{eq:vol-div-raten} (``empirical rate'') and the right-hand side of \eqref{eq:vol-div-raten} (``rate'') versus time. After different initial values mainly due to the differing lengths of the initial interval, the ratios converge. At time 100, the maximal absolute value of all differences between them equals 0.417, at time 1000 it reduces to 0.042. 

The maximum ratio of empirical rate and rate in the previous example at time 1000 equals 3.36, quite a bit away from 1. This is due to the fact that $\vol((f_{0:k}^{-1}\circ\mbf T_C^{-(k+1)})(c(0\!:\!k)))$ grows at rate $\log\lambda$. The reason for this is that the symmetry of the situation allows without loss of generality to assume that the eavesdropper always receives the symbol $c_1$. The volume of states compatible with this sequence is essentially given by the difference of the largest and smallest paths which are possible according to this information, which by Lemma \ref{thm:mpdyn} grows at rate $\log\lambda$. Since the extreme paths compatible with a given eavesdropper information always diverge at rate $\log\lambda$ by Lemma \ref{lem:divcoeff}, a smaller volume growth rate is only possible if there are gaps in the set of possible states, as occur in the case $\Omega=0$ (see above). We expect these gaps to increase if the difference between $\Gamma$ and $R$ increases.

A major problem for the general analysis of $\vol((f_{0:k}^{-1}\circ\mbf T_C^{-t_k})(c(0\!:\!t_k-1)))$ is that a brute-force approach quickly becomes infeasible because with every secure transmission scheme, at least $2^{t+1}$ different message sequences are possible at time $t$ from the eavesdropper's point of view. A general analysis without relying on symmetry might require techniques from fractal set theory. Symmetry as in the example above is simpler to analyze. To achieve this symmetry, the association of quantizer messages to the code sets is crucial, an issue we have neglected here. We also expect the gap between the left- and the right-hand side of \eqref{eq:vol-div-raten} to decrease at higher blocklengths, not least because the $\varepsilon_n$ term in the lower bound at blocklength $n=1$ and with $M=3,\lambda=1.2$ as in the example equals $1.22$ and is not negligible.

\section{Conclusion}\label{sect:concl}

In this paper we introduced uncertain wiretap channels and their zero-error secrecy capacity. We introduced methods from hypergraph theory which together with the already established graph theoretic methods for the zero-error capacity of uncertain channels facilitate the analysis of zero-error secrecy capacity. We showed how the zero-error secrecy capacity of an uncertain wiretap channel relates to the zero-error capacity of the uncertain channel to the intended receiver of the wiretap channel. In the case that the uncertain channel to the intended receiver is injective, we gave a full characterization of the zero-error secrecy capacity of the corresponding uncertain wiretap channel.

We also analyzed how unstable linear systems can be estimated if the system state information has to be transmitted to the estimator through an uncertain wiretap channel, such that the eavesdropper should obtain as little information about the system states as possible. We introduced two security criteria, called d-security and v-security. We gave a sufficient criterion which uncertain channels have to satisfy in order for the estimator to obtain a bounded estimation error as well as both d- and v-security to hold. In the case of an injective uncertain channel from encoder to estimator, we showed that this sufficient criterion essentially is necessary as well. We gave lower bounds on the exponential rates at which the eavesdropper's state information diverges under the two security criteria.

Some problems have been left open in the paper, like a complete characterization of the zero-error secrecy capacity of uncertain wiretap channels, a characterization of when secure estimation of unstable systems is possible over uncertain wiretap channels and a complete answer to the question of optimality of the lower bounds from Lemma \ref{lem:volrate}. Apart from that, there are several points where the paper could be extended in the future. One would be that the encoder has less knowledge about the uncertain wiretap channel. Another one would be an extension to multi-dimensional secure estimation, possibly with distributed observations. Finally, it would be interesting to link the zero-error secrecy capacity of uncertain wiretap channels to Nair's nonstochastic information theory \cite{Nnst} (cf.~Footnote \ref{fn:taxicab}).

\appendices

\section{Uncertain Wiretap Channels: Proofs and Further Discussion}\label{app:channels}

This appendix contains the proofs of Theorems \ref{thm:zescap} and \ref{thm:injch} and sone additional discussion. First we prove Theorem \ref{thm:zescap}. For the proof of Theorem \ref{thm:injch} we then introduce a graph and a hypergraph structure on the input alphabet induced by the uncertain wiretap channel. Using these structures, we prove Theorem \ref{thm:injch}.

\subsubsection{Proof of Theorem \ref{thm:zescap}}\label{app:zescap-proof}

Assume that $C_0(\mbf T_B,\mbf T_C)>0$, which implies $C_0(\mbf T_B)>0$. Let $\mbf F$ be a zero-error wiretap $(n_1,M_1)$-code and let $\mbf G$ be a zero-error $(n_2,M_2)$-code, where $M_1=N_{(\mbf T_B,\mbf T_C)}(n_1)\geq 2$ and $M_2=N_{\mbf T_B}(n_2)$. Consider the concatenated $(n_1+n_2,M_1M_2)$-code $\mbf F\times\mbf G$. Clearly, it is a zero-error code. But it also is a zero-error wiretap code: Choose $(m_1,m_2)\in\{0,\ldots,M_1-1\}\times\{0,\ldots,M_2-1\}$ and choose $c(1\!:\!n_1)\in\mbf T_C^{n_1}(\mbf F(m_1))$ and $c(n_1+1\!:\!n_2)\in\mbf T_C^{n_2}(\mbf G(m_2))$. Since $\mbf F$ is a zero-error wiretap code, there exists an $m_1'\in(\mbf F^{-1}\circ\mbf T_C^{-n_1})(c(1\!:\!n_1))$ with $m_1'\neq m_1$. Therefore the two different message pairs $(m_1,m_2),(m_1',m_2)$ both can generate the output $c(1\!:\!n_1+n_2)$. Thus $\mbf F\times\mbf G$ is a zero-error wiretap code. This construction implies 
\[
	\frac{\log N_{(\mbf T_B,\mbf T_C)}(n_1\!+\!n_2)}{n_1+n_2}\!\geq\!\frac{\log N_{(\mbf T_B,\mbf T_C)}(n_1)\!+\!\log N_{\mbf T_B}(n_2)}{n_1+n_2},
\]
and the term on the right-hand side tends to $C_0(\mbf T_B)$ as $n_2$ tends to infinity. This proves Theorem \ref{thm:zescap}.

\subsubsection{Zero-Error Capacity and Graphs}

It was observed by Shannon \cite{Shad} that the zero-error capacity of an uncertain channel $\mbf T:\mc A\rightarrow2^{\mc B}_*$ can be determined from a graph structure induced on the input alphabet $\mc A$ by $\mbf T$. To see this, let $n$ be a blocklength. Two words $a(1\!:\!n),a'(1\!:\!n)\in\mc A^n$ cannot be used as codewords for the same message if they have a common output word $b(1\!:\!n)\in\mc B^n$. If we draw a line between every two elements of $\mc A^n$ which generate a common output message $b(1\!:\!n)$, we obtain a \textit{graph} on $\mc A^n$ which we denote by $G(\mbf T^n)$. Thus $G(\mbf T^n)$ is nothing but a binary relation $\sim$ on $\mc A^n$, where $a(1\!:\!n)\sim a'(1\!:\!n)$ if and only if $\mbf T^n(a(1\!:\!n))\cap\mbf T^n(a(1\!:\!n))\neq\varnothing$. Since the blocklength should always be clear from the context, we omit it in the $\sim$-notation.

We call a family $\{\mbf F(0),\ldots\mbf F(M-1)\}$ of disjoint subsets of $\mc A^n$ an \textit{independent system in $G(\mbf T^n)$} if for all $m,m'\in\{0,\ldots,M-1\}$ with $m\neq m'$, we have $a(1\!:\!n)\not\sim a'(1\!:\!n)$ for all $a(1\!:\!n)\in\mbf F(m),a'(1\!:\!n)\in\mbf F(m')$. Clearly, every independent system consisting of $M$ disjoint subsets of $\mc A$ is a zero-error $(n,M)$-code for $\mbf T$ and vice versa. Finding the zero-error capacity of $\mbf T$ therefore amounts to finding the asymptotic behavior as $n\rightarrow\infty$ of the sizes of maximum independent systems of the graphs $G(\mbf T^n)$.

Given two blocklengths $n_1,n_2$ and elements $a(1\!:\!n_1+n_2),a'(1\!:\!n_1+n_2)$ of $\mc A^{n_1+n_2}$, note that $a(1\!:\!n_1+n_2)\sim a'(1\!:\!n_1+n_2)$ if and only if one of the following holds:
\begin{enumerate}
  \item $a(1\!:\!n_1)\!=\!a'(1\!:\!n_1)$ and $a(n_1\!+\!1\!:\!n_2)\!\sim\! a'(n_1\!+\!1\!:\!n_2)$,
  \item $a(1\!:\!n_1)\!\sim\! a'(1\!:\!n_1)$ and $a(n_1\!+\!1\!:\!n_2)\!=\!a'(n_1\!+\!1\!:\!n_2)$,
  \item $a(1\!:\!n_1)\!\sim\! a'(1\!:\!n_1)$ and $a(n_1\!+\!1\!:\!n_2)\!\sim\! a'(n_1\!+\!1\!:\!n_2)$.
\end{enumerate}
We can therefore say that $G(\mbf T^{n_1+n_2})$ is the \textit{strong graph product} of $G(\mbf T^{n_1})$ and $G(\mbf T^{n_2})$, see \cite[Definition 1.9.4]{BRgraphtext}. In particular, $G(\mbf T^n)$ is the $n$-fold product of $G(\mbf T)$ with itself.

\subsubsection{Zero-Error Secrecy Capacity and Hypergraphs}

Let $(\mbf T_B,\mbf T_C)$ be an uncertain wiretap channel and $n$ a blocklength. In order to use the above graph-theoretic framework for zero-error capacity also in the treatment of the zero-error secrecy capacity of $(\mbf T_B,\mbf T_C)$, we introduce an additional structure on $\mc A^n$, which is induced by $\mbf T_C$. Every output $c(1\!:\!n)$ of $\mbf T_C^n$ generates the set $e\e n(c(1\!:\!n)):=\mbf T_C^{-n}(c(1\!:\!n))\subseteq\mc A^n$. We set $\mc E(\mbf T_C^n):=\{e\e n(c(1\!:\!n)):c(1\!:\!n)\in\ran(\mbf T_C^n)\}$. Every element $e\e n$of $\mc E(\mbf T_C^n)$ is called a \textit{hyperedge} and the pair $(\mc A^n,\mc E(\mbf T_C^n))$ a \textit{hypergraph} denoted by $H(\mbf T_C^n)$.

Now let $\mbf F$ be a zero-error $(n,M)$-code for $\mbf T_B$. Then by definition, it is a zero-error wiretap $(n,M)$-code for $(\mbf T_B,\mbf T_C)$ if and only if $\card\{m:\mbf F(m)\cap e\e n\}\geq 2$ for every $e\e n\in\mc E(\mbf T_C^n)$. In other words, together with the above observation about zero-error codes and graphs we obtain the following lemma.

\begin{lem}\label{lem:zecrit}
  A family $\{\mbf F(0),\ldots,\mbf F(M-1)\}$ of disjoint subsets of $\mc A^n$ is a zero-error wiretap $(n,M)$-code for $(\mbf T_B,\mbf T_C)$ if and only if it is an independent system in $G(\mbf T_B^n)$ and if $\card\{m:\mbf F(m)\cap e\e n\}\geq 2$ for every $e\e n\in\mc E(\mbf T_C^n)$.
\end{lem}

\begin{figure}
	\centering
	\subfloat[]{
	\begin{tikzpicture}[vertex/.style={draw,circle,fill=black,minimum size=4pt, inner sep=0pt}, scale=0.8, every node/.style={transform shape}]%
	\draw (0,0) node[vertex] (x1) {}
		 ++(0,1.4) node[vertex] (x2) {}
		-- ++(1.4,0) node[vertex] (x3) {}
		 ++(0,-1.4) node[vertex] (x4) {};
		 
	\draw (0,.2) node {$a_1$}
		++ (0,.9) node {$a_2$}
		++ (1.4,0) node {$a_3$}
		++ (0,-.9) node {$a_4$};

	\draw [blue, dotted, thick] plot [smooth cycle, tension = 0.8] coordinates {(0,-.3) (.5,.7) (0,1.7) (-.5,.7)};
	\draw [blue, dotted, thick] plot [smooth cycle, tension = 0.8] coordinates {(1.4,-.3) (1.9,.7) (1.4,1.7) (.9,.7)};
\end{tikzpicture}
}
\hfil
\subfloat[]{
\begin{tikzpicture}[every node/.style={draw,circle,fill=white,minimum size=8pt, inner sep=1pt, transform shape}, scale=0.8]%
	\draw (0,0) node (x1) {\footnotesize 0}
		 ++(0,1.4) node (x2) {\footnotesize 1}
		-- ++(1.4,0) node (x3) {\footnotesize 1}
		 ++(0,-1.4) node (x4) {\footnotesize 2};

	\draw [blue, dotted, thick] plot [smooth cycle, tension = 0.8] coordinates {(0,-.3) (.5,.7) (0,1.7) (-.5,.7)};
	\draw [blue, dotted, thick] plot [smooth cycle, tension = 0.8] coordinates {(1.4,-.3) (1.9,.7) (1.4,1.7) (.9,.7)};
\end{tikzpicture}
}
\caption{(a): The pair $(G(\mbf T_B),H(\mbf T_C))$ corresponding to the uncertain wiretap channel $(\mbf T_B,\mbf T_C)$ from Fig.~\ref{fig:uncch-code}(b). The black, solid line means that $a_2$ and $a_3$ are adjacent to each other in $G(\mbf T_B)$. The blue, dotted lines are the boundaries of the hyperdeges of $H(\mbf T_C)$. (b): The number inscribed in each node indicates to which set $\mbf F(m)$ the node belongs, where $\mbf F$ is the zero-error wiretap code defined in Fig.~\ref{fig:uncch-code}(b).}\label{fig:grhyp}
\end{figure}
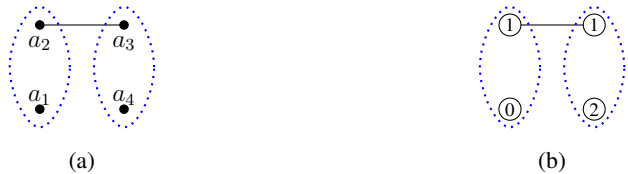

Observe that every $e\e n\in\mc E(\mbf T_C^n)$ has the form $e_1\times\cdots\times e_n$ for some $e_1,\ldots,e_n\in\mc E(\mbf T_C)$, and that every Cartesian product $e_1\times\cdots\times e_n$ of elements of $\mc E(\mbf T_C)$ is an element of $\mc E(\mbf T_C^n)$. This means that $H(\mbf T_C^n)$ is the \textit{square product} of $H(\mbf T_C)$ (see \cite{HOShyperprod}). For the uncertain wiretap channel from Fig. \ref{fig:uncch-code}(b), the corresponding graph/hypergraph pair at blocklength 1 and a zero-error wiretap code are illustrated in Fig. \ref{fig:grhyp}.

\subsubsection{Proving Theorem \ref{thm:injch}}\label{app:injch-proof}

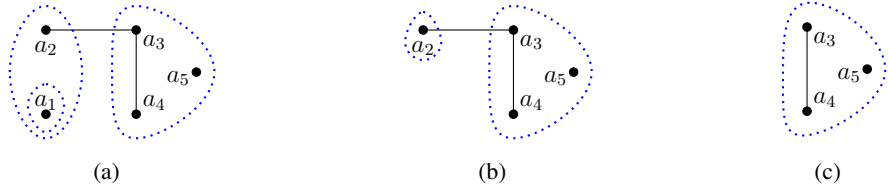
\begin{figure}[!t]
	\centering
	\subfloat[]{
	\begin{tikzpicture}[vertex/.style={draw,circle,fill=black,minimum size=4pt, inner sep=0pt}, scale=0.8, every node/.style={transform shape}]
		\draw (0,0) node[vertex] (x1) {}
		 ++(0,1.4) node[vertex] (x2) {}
		-- ++(1.5,0) node[vertex] (x3) {}
		-- ++(0,-1.4) node[vertex] (x4) {}
		 ++ (1,.7) node[vertex] (x5) {};
		 
		 \draw (0,.2) node (a1) {$a_1$}
		 ++ (0,.9) node (a2) {$a_2$}
		 ++ (1.8,0.1) node (a3) {$a_3$}
		 ++ (0,-1) node (a4) {$a_4$}
		 ++ (.4,.4) node (a5) {$a_5$};
		 
		 \draw [blue, dotted, thick] plot [smooth cycle, tension = 0.8] coordinates {(-.3,0.15) (0,.5) (.3,0.15) (0,-.3)};
		 \draw [blue, dotted, thick] plot [smooth cycle, tension = 0.9] coordinates {(0,-.4) (-.6,.7) (0,1.8) (.6,.7)};
		 \draw [blue, dotted, thick] plot [smooth cycle, tension = 0.9] coordinates {(1.5,-.4) (1.1,.7) (1.5,1.8) (2.8,.7)};
	\end{tikzpicture}}
	\hfil
	\subfloat[]{
	\begin{tikzpicture}[vertex/.style={draw,circle,fill=black,minimum size=4pt, inner sep=0pt}, scale=0.8, every node/.style={transform shape}]
		\draw (0,1.4) node[vertex] (x2) {}
		-- ++(1.5,0) node[vertex] (x3) {}
		-- ++(0,-1.4) node[vertex] (x4) {}
		 ++ (1,.7) node[vertex] (x5) {};
		 
		 \draw (0,1.1) node (a2) {$a_2$}
		 ++ (1.8,0.1) node (a3) {$a_3$}
		 ++ (0,-1) node (a4) {$a_4$}
		 ++ (.4,.4) node (a5) {$a_5$};
		 
		 \draw [blue, dotted, thick] plot [smooth cycle, tension = 0.8] coordinates {(-.3,1.2) (0,1.7) (.3,1.2) (0,.9)};
		 \draw [blue, dotted, thick] plot [smooth cycle, tension = 0.9] coordinates {(1.5,-.4) (1.1,.7) (1.5,1.8) (2.8,.7)};
	\end{tikzpicture}}
	\hfil
	\subfloat[]{
	\begin{tikzpicture}[vertex/.style={draw,circle,fill=black,minimum size=4pt, inner sep=0pt}, scale=0.8, every node/.style={transform shape}]
		\draw (1.4,0) node[vertex] (x3) {}
		-- ++(0,-1.4) node[vertex] (x4) {}
		 ++ (1,.7) node[vertex] (x5) {};
		 
		 \draw (1.7,-0.2) node (a3) {$a_3$}
		 ++ (0,-1) node (a4) {$a_4$}
		 ++ (.4,.4) node (a5) {$a_5$};
		 
		 \draw [blue, dotted, thick] plot [smooth cycle, tension = 0.9] coordinates {(1.4,-1.8) (1,-.9) (1.4,.4) (2.7,-.7)};
	\end{tikzpicture}}
	\caption{(a): The original graph/hypergraph pair $(G(\mbf T_B),H(\mbf T_C))$ of some uncertain wiretap channel $(\mbf T_B,\mbf T_C)$. $a_1$ cannot be used in any zero-error wiretap code. (b): If $a_1$ is not used in any zero-error wiretap code, then $a_2$ is unusable as well. (c): Having eliminated $a_1$ and $a_2$, there are no singletons or cliques left among the hyperedges.}\label{fig:elim}
\end{figure}

Theorem \ref{thm:injch} will follow from a slightly more general lemma which holds for general wiretap channels. This lemma analyzes a procedure, to be presented next, which eliminates elements $a(1\!:\!n)$ from $\mc A^n$ which do not satisfy a necessary condition for being a codeword of a zero-error wiretap code. The idea behind the procedure is that by Lemma \ref{lem:zecrit} no $a(1\!:\!n)\in\mc A^n$ can be a codeword which is contained in an $e\e n\in\mc E(\mbf T_C^n)$ which is a singleton or where all elements of $e\e n$ are connected in $G(\mbf T_B^n)$. Thus these elements can be neglected when looking for a zero-error wiretap code. This amounts to deleting those elements from the input alphabet and to restricting the wiretap channel to the reduced alphabet. But not using a certain subset of the input alphabet may generate yet another set of unusable input words. Thus a further reduction of the input alphabet may be necessary, and so on, see Fig. \ref{fig:elim}. We now formalize this procedure and analyze the result.

We apply the graph/hypergraph language developed above and start with introducing some related terminology. Let $(\mbf T_B,\mbf T_C)$ be an uncertain wiretap channel with input alphabet $\mc A$. For any subset $\mc A'$ of $\mc A$, one can consider the uncertain wiretap channel restricted to inputs from $\mc A'$, thus creating an uncertain wiretap channel $(\mbf T_B\vert_{\mc A'}:\mc A'\rightarrow2^{\mc B}_*,\mbf T_C\vert_{\mc A'}:\mc A'\rightarrow2^{\mc C}_*)$ satisfying $\mbf T_B\vert_{\mc A'}(a)=\mbf T_B(a)$ and $\mbf T_C\vert_{\mc A'}(a)=\mbf T_C(a)$ for all $a\in\mc A'$. Thus, $\mbf T_B\vert_{\mc A'}$ generates a graph $G(\mbf T_B\vert_{\mc A'})$ on $\mc A'$ and $\mbf T_C\vert_{\mc A'}$ generates a hypergraph $H(\mbf T_C\vert_{\mc A'})$ on $\mc A'$. If we say that we \textit{eliminate} a set $\mc V$ from $G(\mbf T_B)$ or $H(\mbf T_C)$, we mean that we pass from $G(\mbf T_B)$ to $G(\mbf T_B\vert_{\mc A\setminus\mc V})$ or from $H(\mbf T_C)$ to $H(\mbf T_C\vert_{\mc A\setminus\mc V})$, respectively. Further, a \textit{clique in $G(\mbf T_B)$} is a subset $\mc V\subseteq\mc A$ such that $a\sim a'$ for all $a,a'\in\mc V$. We write 
\begin{align*}
  &\mc E(G(\mbf T_B),H(\mbf T_C))_{s,c}\\
  &:=\{e\in\mc E(\mbf T_C):\card e=1\text{ or }e\text{ is clique in }G(\mbf T_B)\}.
\end{align*}

Finally, we can formalize the procedure of deleting some of the unusable input words from the input alphabet of an uncertain wiretap channel. Let $(\mbf T_B,\mbf T_C)$ be an uncertain wiretap channel with input alphabet $\mc A$ and fix a blocklength $n\geq 1$. For the sake of shorter notation, we use the notation $a^n$ for elements of $\mc A^n$ in the rest of the section. Put $\mc A_{s,c}\e n(-1)=\varnothing$ and for $i\geq 0$ set
\begin{align}
  &\!\!G\e n(i):=G(\mbf T_B^n\vert_{\mc A^n\setminus\mc A_{s,c}\e n(i-1)}),\label{eq:graph-gen}\\
  &\!\!H\e n(i):=H(\mbf T_C^n\vert_{\mc A^n\setminus\mc A_{s,c}\e n(i-1)}),\label{eq:hyper-gen}\\
  &\!\!\mc A_{s,c}\e n(i)\!:=\!\{a^n\!\!:\exists\, e\e n\!\!\in\!\mc E(G\e n\!(i),\!H\e n\!(i))_{s,c}:a^n\in e\e n\}\notag\\&\qquad\qquad\cup\mc A_{s,c}\e n(i-1).\label{eq:set-red}
\end{align}
Note that $\mc A_{s,c}\e n(-1)\subseteq\mc A_{s,c}\e n(0)\subseteq\mc A_{s,c}\e n(1)\subseteq\cdots$. Define 
\begin{align*}
 I\e n&:=[\min\{i\geq -1:\mc A_{s,c}\e n(i+1)=\mc A_{s,c}\e n(i)\}]_+,\\
 \mc A_{s,c}\e n&:=\mc A_{s,c}\e n(I\e n)
\end{align*}
where we set $[x]_+=\max\{x,0\}$ for any real number $x$. Thus $I\e n+1$ is the number of steps of the procedure \eqref{eq:graph-gen}-\eqref{eq:set-red} where the input alphabet is strictly reduced. The reason for defining $I\e n$ in the way we have done will become clear in the proof of Lemma \ref{lem:elim-square} below. Since $\mc A$ is finite, clearly $I\e n<\infty$.

The next lemma says that not being an element of $\mc A_{s,c}\e n$ is a necessary condition for any $a^n\in\mc A^n$ to be the codeword of a zero-error wiretap code.

\begin{lem}
  If $\mbf F$ is a zero-error wiretap $(n,M)$-code for $(\mbf T_B,\mbf T_C)$ and any $M\geq 2$, then $\ran(\mbf F)\cap\mc A_{s,c}\e n=\varnothing$.
\end{lem}

\begin{IEEEproof}
  We use induction over the reduction steps $i$. Let $M\geq 2$ and assume that $\mbf F$ is a zero-error wiretap $(n,M)$-code for $(\mbf T_B,\mbf T_C)$. By Lemma \ref{lem:zecrit} it is clear that $\ran(\mbf F)\cap\mc A_{s,c}\e n(0)=\varnothing$. Thus $\mbf F$ also is a zero-error wiretap $M$-code for the reduced uncertain wiretap channel $(\mbf T_B^n\vert_{\mc A^n\setminus\mc A_{s,c}\e n(0)},\mbf T_C^n\vert_{\mc A^n\setminus\mc A_{s,c}\e n(0)})$. In particular, if $e\e n\in\mc E(G\e n(1),H\e n(1))_{s,c}$, then $e\e n\cap\ran(\mbf F)=\varnothing$. Now note that the union of all $e\e n\in\mc E(G\e n(1),H\e n(1))_{s,c}$ equals $\mc A_{s,c}\e n(1)\setminus\mc A_{s,c}\e n(0)$. Therefore $\ran(\mbf F)\cap\mc A_{s,c}\e n(1)=\varnothing$. Repeating this argument $I\e n$ times, one obtains the statement of the lemma.
\end{IEEEproof}

The crucial point about the above elimination procedure is that one can relate $\mc A_{s,c}\e n$ to $\mc A_{s,c}\e 1$, which in turn will give us Theorem \ref{thm:injch}.

\begin{lem}\label{lem:elim-square}
  For any uncertain wiretap channel $(\mbf T_B,\mbf T_C)$ and every blocklength $n\geq 1$, the corresponding set $\mc A_{s,c}\e n$ satisfies $\mc A_{s,c}\e n=(\mc A_{s,c}\e 1)^n$. 
\end{lem}

Before proving Lemma \ref{lem:elim-square}, we show how Theorem \ref{thm:injch} follows from it.

\begin{IEEEproof}[Proof of Theorem \ref{thm:injch}]
Observe that one can restrict attention to singleton zero-error wiretap codes because the injectivity of $\mbf T_B$ implies that no vertices are connected in $G(\mbf T_B^n)$ for any $n$. Further, since $H(\mbf T_C\vert_{\mc A^n\setminus\mc A_{s,c}\e n})$ has no singletons as hyperedges by construction of $\mc A_{s,c}\e n$, we conclude that $N_{(\mbf T_B,\mbf T_C)}(n)=(\card\mc A)^n-\card\mc A_{s,c}\e n$. By Lemma \ref{lem:elim-square}, we have $\card\mc A_{s,c}\e n=(\card\mc A_{s,c}\e 1)^n$. Thus if $\mc A_{s,c}\e1$ is a strict subset of $\mc A$, then 
\[
	C_0(\mbf T_B,\mbf T_C)=\lim_{n\rightarrow\infty}\frac{\log N_{(\mbf T_B,\mbf T_C)}(n)}{n}=\log\card\mc A.
\]
Otherwise, $C_0(\mbf T_B,\mbf T_C)$ obviously equals 0. This proves Theorem \ref{thm:injch}.
\end{IEEEproof}

\begin{IEEEproof}[Proof of Lemma \ref{lem:elim-square}] 
 Fix $n\geq 2$. We set $\sigma:=I\e 1$ and define a mapping $\iota:\mc A\rightarrow\{0,\ldots,\sigma\}\cup\{\infty\}$,
\[
	\iota(a)=\begin{cases}
	          \text{the }i\text{ with }a\in\A{i}\setminus\A{i-1}&\text{if }a\in\mc A_{s,c}\e n,\\
	          \infty & \text{otherwise}.
	         \end{cases}
\]
We also define 
\[
  \iota\e n(a^n)=(\iota(a_1),\ldots,\iota(a_n)).
\]
Similarly, for $e\in\mc E(\mbf T_C)$ with $e\subset\mc A_{s,c}\e 1$ we set $\iota(e):=\max\{\iota(a):a\in e\}$, and for any $e\e n\in\mc E(\mbf T_C^n)$ we define $\iota\e n(e\e n)=(\iota(e_1),\ldots,\iota(e_n))$. 

For any $i^n\in(\{0,\ldots,\sigma\}\cup\{\infty\})^n$ we set 
\begin{align*}
	f(i^n)=\{a^n\in(\mc A_{s,c}\e 1)^n:\iota\e n(a^n)=i^n\},\quad
	w(i^n)=\sum_{t=1}^ni_t
\end{align*}
and for $0\leq\mu\leq n\sigma$
\[
	F(\mu):=\bigcup_{i^n\in\{0,\ldots,\sigma\}^n:w(i^n)\leq\mu}f(i^n).
\]
Note that $F(n\sigma)=(\mc A_{s,c}\e 1)^n$. We will now prove 
\begin{align}
	F(\mu)&=\A[n]{\mu} \quad\text{for }0\leq\mu\leq n\sigma,\label{eq:zz}\\
	I\e n&=n\sigma=nI\e 1.\label{eq:azz}
\end{align}
Together, \eqref{eq:zz} and \eqref{eq:azz} imply $(\mc A_{s,c}\e 1)^n=F(nI\e 1)=\mc A_{s,c}\e n$, which is what we want to prove.

We first prove \eqref{eq:zz} by induction over $\mu$. Let $\mu=0$. Then $F(0)=(\mc A_{s,c}\e 1(0))^n$. This is easily seen to equal $\mc A_{s,c}\e n(0)$.

Next let $0\leq\mu\leq n\sigma-1$ and assume \eqref{eq:zz} has been proven for all $0\leq\mu'\leq\mu$. We need to show that \eqref{eq:zz} holds for $\mu+1$. First we show that $F(\mu+1)\subseteq\A[n]{\mu+1}$.

Let $i^n\in\{0,\ldots,\sigma\}^n$ with $w(i^n)=\mu+1$. We have to show that $f(i^n)\subseteq\A[n]{\mu+1}$. Choose an $a^n$ with $\iota(a^n)=i^n$. Then by \eqref{eq:set-red}, for every $1\leq t\leq n$, there exists an $e_t\in\mc E(\mbf T_C)$ such that $a^n\in e\e n=e_1\times\cdots\times e_n$ and $\iota\e n(e\e n)=i^n$. Therefore
\begin{align}\label{eq:endlich}
	\!\!\!\!\!\!\!&e\e n\setminus\A[n]{\mu}\stackrel{(a)}{=}e\e n\setminus F(\mu)\notag\\
	&\stackrel{(b)}{=}(e_1\!\setminus\!\A{\iota(e_1)\!-\!1})\!\times\!\cdots\!\times\!(e_n\!\setminus\!\A{\iota(e_n)\!-\!1}),
\end{align}
where $(a)$ is due to the induction hypothesis and $(b)$ holds because $e_t\setminus\mc A_{s,c}\e 1(\iota(e_t))=\varnothing$.
By definition of the mapping $\iota$, every set $e_t\setminus\A{\iota(e_t)-1}$ is a singleton or a clique, hence so is the right-hand side of \eqref{eq:endlich}. Thus $e\e n\setminus\A[n]{\mu}\in\mc E(G\e n(\mu+1),H\e n(\mu+1))_{s,c}$, hence $a^n\in\A[n]{\mu+1}$. This proves $F(\mu+1)\subseteq\mc A_{s,c}\e n(\mu+1)$.

Now we prove that $\A[n]{\mu+1}\subseteq F(\mu+1)$, which is equivalent to showing that $\mc A^n\setminus F(\mu+1)\subseteq\mc A^n\setminus\A[n]{\mu+1}$. Let $a^n\in\mc A^n\setminus F(\mu+1)$. Thus $a^n\in\mc A^n\setminus F(\mu)=\mc A^n\setminus\mc A_{s,c}\e n(\mu)$, where the equality is due to the induction hypothesis. We need to show that $e\e n\setminus\mc A_{s,c}\e n(\mu)\not\subseteq\mc A_{s,c}\e n(\mu+1)$ for every $e\e n\in\mc E(\mbf T_C^n)$ containing $a^n$, since then $a^n\in\mc A^n\setminus\mc A_{s,c}\e n(\mu+1)$. 
 
Choose any $e\e n=e_1\times\cdots\times e_n\in\mc E(\mbf T_C^n)$ containing $a^n$. Let $i^n=\iota\e n(a^n)$. Thus $\iota(e_t)\geq i_t$ for every $t\in\{1,\ldots,n\}$. Choose any $t_*\in\{1,\ldots,n\}$. If $0\leq i_{t_*}\leq \sigma$, there exists an $a_{t_*}'\in e_{t_*}\setminus\mc A_{s,c}\e 1(i_{t_*}-2)$ with $a_{t_*}\not\sim a_{t_*}'$ because otherwise, $e_t\setminus\mc A_{s,c}\e 1(i_{t_*}-2)$ would be a singleton or a clique in $G\e 1(i_{t_*}-1)$, hence a subset of $\mc A_{s,c}\e 1(i_{t_*}-1)$, which we know not to be true because $\iota(e_{t_*})\geq i_{t_*}$. A similar argument shows that there exists an $a'_{t_*}\in\mc A\setminus\mc A_{s,c}\e 1$ with $a_{t_*}\not\sim a_{t_*}$ if $i_{t_*}=\infty$. Consequenctly, the sequence $\tilde a^n=(a_1,\ldots,a_{t_*-1},a_{t_*}',a_{t_*+1},\ldots,a_n)$ is an element of $e\e n$ satisfying $a^n\not\sim\tilde a^n$ because $G(\mbf T_B^n)$ is the $n$-fold strong graph product of $G(\mbf T_B)$. Notice that $w(\tilde a^n)\geq w(a^n)-1\geq\mu+1$ because $\iota(a_{t_*}')\geq i_{t_{*}}-1$. In particular, $\tilde a^n\notin F(\mu)=\mc A_{s,c}\e n(\mu)$. Thus we have found two different $a^n,\tilde a^n\in e\e n\setminus\mc A_{s,c}\e n(\mu)$ which are not adjacent to each other, which implies $e\e n\setminus\mc A_{s,c}\e n(\mu)\not\subseteq\mc A_{s,c}\e n(\mu+1)$. This is what we had to prove to show that $\mc A^n\setminus F(\mu+1)\subseteq\mc A^n\setminus\mc A_{s,c}\e n(\mu+1)$, and this completes the proof of \eqref{eq:zz}.

To show \eqref{eq:azz}, observe that \eqref{eq:zz} implies $nI\e 1\leq I\e n$. If $I\e n>nI\e1=n\sigma$, then\linebreak $\mc E(G\e n(n\sigma),H\e n(n\sigma))_{s,c}\neq\varnothing$, i.e., there exists an $e\e n=e_1\times\cdots\times e_n\in\mc E(\mbf T_C^n)$ such that $e\e n\setminus\mc A_{s,c}\e n(n\sigma)=e\e n\setminus F(n\sigma)$ is a clique or a singleton. But if $e\e n\setminus F(n\sigma)\neq\varnothing$, there exists a $t$ such that $e_t\setminus\mc A_{s,c}\e 1\neq\varnothing$, which implies the existence of $a_t,a_t'\in e_t\setminus\mc A_{s,c}\e 1$ with $a_t\neq a_t'$ and $a_t\not\sim a_t'$. If the $t$-th component of $a^n\in e\e n$ equals $a_t$, then $a^n\notin F(n\sigma)=\mc A_{s,c}\e n(n\sigma)$. But then $\tilde a^n:=(a_1,\ldots,a_{t-1},a_t',a_{t+1},\ldots,a_n)\in e\e n\setminus\mc A_{s,c}\e n(n\sigma)$ as well. Since $a^n\not\sim\tilde a^n$, this implies that $e\e n\setminus\mc A_{s,c}\e n(n\sigma)$ is neither a clique nor a singleton. Thus $I\e n=nI\e 1$, which proves \eqref{eq:azz}. This completes the proof of Theorem \ref{thm:injch}.
\end{IEEEproof}

\subsubsection{Examples and Discussion}\label{app:exdiscres}

\begin{example}\label{ex:singletonnec}
	The uncertain wiretap channel $(\mbf T_B,\mbf T_C)$ shown in Fig.\ \ref{fig:uncch-code}(b) is an example of the fact that at finite blocklengths $n$, non-singleton zero-error wiretap codes may be necessary to achieve $N_{(\mbf T_B,\mbf T_C)}(n)$. If one applies the zero-error wiretap code $\mbf F=\{\{a_1\},\{a_2,a_3\},\{a_4\}\}$, then three messages can be distinguished at the intended receiver's output and every eavesdropper output can be generated by two different messages. Hence $\mbf F$ is a zero-error wiretap $(1,3)$-code. On the other hand, the maximal $M$ for which a singleton zero-error wiretap $(1,M)$-code exists is $M=2$, for example $\mbf F=\{\{a_1\},\{a_4\}\}$. $M=4$ is not possible because $N_{\mbf T_B}(1)=3$. For $M=3$, either $c_1$ or $c_2$ would be generated by only one message. 
	 
	 We conjecture that non-singleton zero-error wiretap codes are also necessary to achieve $C_0(\mbf T_B,\mbf T_C)$.
\end{example}

One can also construct examples which show the following: If there exists a zero-error wiretap $(n,M)$-code, then it is necessary to have non-singleton codes to also find a zero-error wiretap $(M',n)$-code for every $2\leq M'\leq M$. 

Another open question is when the zero-error wiretap capacity of general uncertain wiretap channels is positive.

\begin{example}\label{ex:superact}

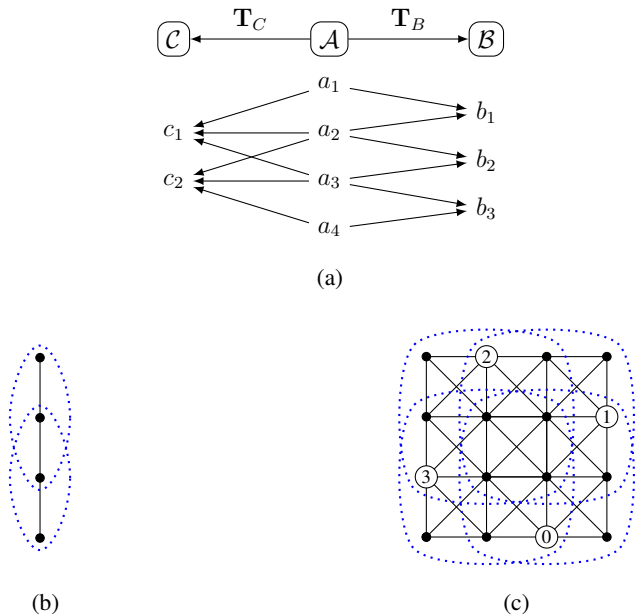
\begin{figure}
\centering
    \subfloat[]{
	\begin{tikzpicture}[alphabet/.style={draw, rounded corners}, pfeil/.style={->, >=latex}, scale=0.8, every node/.style={transform shape}]
		\node[alphabet] (Eve) {$\mc C$};
		\node[alphabet, right = 2cm of Eve] (Alice) {$\mc A$};
		\node[alphabet, right = 2cm of Alice] (Bob) {$\mc B$};
		
		\node[below = .2cm of Alice] (a1) {$a_1$};
		\node[below = 1cm of Alice] (a2) {$a_2$};
		\node[below = 1.8cm of Alice] (a3) {$a_3$};
		\node[below = 2.6cm of Alice] (a4) {$a_4$};
		
		\node[below = 1cm of Eve] (c1) {$c_1$};
		\node[below = 1.8cm of Eve] (c2) {$c_2$};
		
		\node[below = .6cm of Bob] (b1) {$b_1$};
		\node[below = 1.4cm of Bob] (b2) {$b_2$};
		\node[below = 2.2cm of Bob] (b3) {$b_3$};
		
		\draw[pfeil] (Alice) -> (Bob) node[midway, above] {$\mbf T_B$};
		\draw[pfeil] (Alice) -> (Eve) node[midway, above] {$\mbf T_C$};
		
		\draw[pfeil] (a1) -- (b1);
		\draw[pfeil] (a1) -- (c1);
		\draw[pfeil] (a2) -- (b1);
		\draw[pfeil] (a2) -- (b2);
		\draw[pfeil] (a2) -- (c1);
		\draw[pfeil] (a2) -- (c2);
		\draw[pfeil] (a3) -- (b2);
		\draw[pfeil] (a3) -- (b3);
		\draw[pfeil] (a3) -- (c1);
		\draw[pfeil] (a3) -- (c2);
		\draw[pfeil] (a4) -- (b3);
		\draw[pfeil] (a4) -- (c2);
	\end{tikzpicture}
	}\\
	\subfloat[]{%
\begin{tikzpicture}[every node/.style={draw,circle,fill=black,minimum size=4pt, inner sep=0pt, transform shape}, scale=0.8]%
	\draw (0,0) node (x1) {}
		-- ++(0,1) node (x2) {}
		-- ++(0,1) node (x3) {}
		-- ++(0,1) node (x4) {};
		
	\node[draw=white, fill = white, below=.3cm of x1]  {};
		
	\draw [blue, dotted, thick] plot [smooth cycle, tension = 0.8] coordinates {(0,-.2) (.5,1) (0,2.2) (-.5,1)};
	\draw [blue, dotted, thick] plot [smooth cycle, tension = 0.8] coordinates {(0,.8) (.5,2) (0,3.2) (-.5,2)};
\end{tikzpicture}
}
\hfil
\subfloat[]{
	\begin{tikzpicture}[every node/.style={draw,circle,fill=black,minimum size=4pt, inner sep=0pt, transform shape}, scale=0.8]
	\draw (1,0) -- (1,3);
	\draw (2,0) -- (2,3);
	\draw (0,1) -- (3,1);
	\draw (0,2) -- (3,2);
	
	\draw (0,2) -- (1,3);
	\draw (0,1) -- (2,3);
	\draw (0,0) -- (3,3);
	\draw (1,0) -- (3,2);
	\draw (2,0) -- (3,1);
	
	\draw (0,2) -- (2,0);
	\draw (0,1) -- (1,0);
	\draw (0,3) -- (3,0);
	\draw (1,3) -- (3,1);
	\draw (2,3) -- (3,2);
	
	\draw (0,0) node (00) {} -- ++(1,0) node {} -- ++(1,0) node[draw, fill=white, shape=circle, minimum size=8pt, inner sep=1pt] {\footnotesize 0} -- ++(1,0) node {}
		-- ++(0,1) node {} -- ++(0,1) node[draw,fill = white, shape=circle, minimum size=8pt, inner sep=1pt] {\footnotesize 1} -- ++(0,1) node {}
		-- ++(-1,0) node {} -- ++(-1,0) node[draw, fill = white, shape=circle, minimum size=8pt, inner sep=1pt] {\footnotesize 2} -- ++(-1,0) node{}
		-- ++(0,-1) node {} -- ++(0,-1) node[draw,fill = white, shape=circle, minimum size=8pt, inner sep=1pt] {\footnotesize 3} -- ++(0,-1) node {};
		
	\draw (1,1) node {} -- (2,1) node {} -- (2,2) node {} -- (1,2) node {};

	\draw [blue, dotted, thick] plot [smooth cycle, tension = 0.5] coordinates {(-.2,-.2) (-.2,2.2) (2.2,2.2) (2.2,-.2)};
	\draw [blue, dotted, thick] plot [smooth cycle, tension = 0.5] coordinates {(-.2,.8) (-.2,3.2) (2.2,3.2) (2.2,.8)};
	\draw [blue, dotted, thick] plot [smooth cycle, tension = 0.5] coordinates {(.8,-.2) (.8,2.2) (3.2,2.2) (3.2,-.2)};
	\draw [blue, dotted, thick] plot [smooth cycle, tension = 0.5] coordinates {(.8,.8) (.8,3.2) (3.2,3.2) (3.2,.8)};
\end{tikzpicture}
}
\caption{(a): An uncertain wiretap channel $(\mbf T_B,\mbf T_C)$. (b): $\mc A$ with $G(\mbf T_B)$ and $H(\mbf T_C)$. (c): $\mc A^2$ with $G(\mbf T_B^2)$ and $H(\mbf T_C^2)$. Vertices connected by a solid black line are connected in $G(\mbf T_B)$ or $G(\mbf T_B^2)$, respectively. Vertices within the boundary of a blue dotted line belong to the same hyperedge of $H(\mbf T_C)$ or $H(\mbf T_C^2)$, respectively. A zero-error wiretap $(2,4)$-code is indicated on the right-hand figure.}\label{fig:superact}
\end{figure}

	Consider the wiretap channel $(\mbf T_B,\mbf T_C)$ from Fig.\ \ref{fig:superact}(a). Fig. \ref{fig:superact}(b) shows $\mc A$ with $G(\mbf T_B)$ and $H(\mbf T_C)$ and Fig. \ref{fig:superact}(c) shows $\mc A^2$ with $G(\mbf T_B^2)$ and $H(\mbf T_C^2)$. The code shown in Fig.~\ref{fig:superact}(c) shows that $C_0(\mbf T_B,\mbf T_C)\geq 1$. Since $C_0(\mbf T_B)=1$ by \cite{Shad}, we can even conclude $C_0(\mbf T_B,\mbf T_C)=1$. 
	
	Note that $N_{(\mbf T_B,\mbf T_C)}(1)=1$. Thus the number of messages which can be transmitted securely jumps from none at blocklength 1 to 4 at blocklength 2. This behavior is remarkable when compared to the behavior of zero-error codes for uncertain channels: An uncertain channel $\mbf T$ has $C_0(\mbf T)>0$ if and only if $N_{\mbf T}(1)\geq 2$. This is a simple criterion to decide at blocklength 1 whether or not the zero-error capacity of an uncertain channel is positive. We do not yet have a general simple criterion for deciding whether the zero-error secrecy capacity of an uncertain wiretap channel is positive. Of course, if $\mbf T_B$ is injective, then Theorem \ref{thm:injch} provides such a criterion.
\end{example}

\section{Proofs from Quantizer Analysis}\label{app:quantana}

For reference, we note the following simple lemma which is easily proved by induction.

\begin{lem}\label{lem:induction}
  Let $\mu$ be a real number and let $y(0\!:\!\infty),v(0\!:\!\infty)$ be two sequences of real numbers satisfying $y(t+1)=\mu y(t)+v(t)$ for every $t\geq 0$. Then for every $t\geq 0$
\[
 y(t)=\mu^ty(0)+\sum_{i=0}^{t-1}\mu^{t-i-1}v(i).
\]
\end{lem}

\begin{IEEEproof}[Proof of Lemma \ref{thm:estdyn}]
Note that the quantizer set $\mc P(m(0\!:\!t))$ is an interval. Thus \eqref{eq:A(n),B(n)} implies $\lvert\mc I(m(0\!:\!t+1))\rvert=\lambda\lvert\mc P(m(0\!:\!t))\rvert+\Omega$. Hence by \eqref{eq:Pmn}
	\begin{equation}\label{eq:Pmnlength}
		\lvert\mc P(m(0\!:\!t\!+\!1))\rvert\!=\!\frac{\lvert\mc I(m(0\!:\!t\!+\!1))\rvert}{M}\!=\!\frac{\lambda}{M}\lvert\mc P(m(0\!:\!t))\rvert\!+\!\frac{\Omega}{M}.
	\end{equation}
	Therefore by Lemma \ref{lem:induction}, 
	\begin{align*}
	  \lvert\mc P(m(0\!:\!t))\rvert
	  &=\left(\frac{\lambda}{M}\right)^{t}\lvert\mc P(m(0\!:\!0))\rvert+\frac{\Omega}{M}\sum_{i=0}^{t-1}\left(\frac{\lambda}{M}\right)^{t-i-1}\\
	  &=\left(\frac{\lambda}{M}\right)^t\left(\frac{\lvert I_0\rvert}{M}-\frac{\Omega}{M-\lambda}\right)+\frac{\Omega}{M-\lambda},
	\end{align*}
which proves \eqref{eq:intlength}. The other statements of the lemma are immediate from \eqref{eq:intlength}.
\end{IEEEproof}

\begin{IEEEproof}[Proof of Lemma \ref{lem:nullklar}]
	Let $m(0\!:\!t)\neq m'(0\!:\!t)$. It is sufficient to show that the minimal distance between $\hat x(m(0\!:\!t))$ and $\hat x(m'(0\!:\!t))$ is lower-bounded by $\ell_t$. By Lemma \ref{thm:mpdyn},
\begin{align}\label{eq:Omnull-diff}
	\hat x(m(0\!:\!t))-\hat x(m'(0\!:\!t))
	=\lambda^t\frac{\lvert\mc I_0\rvert}{M}\underbrace{\sum_{i=0}^t\frac{m(i)-m'(i)}{M^{i}}}_{=:n(m,m',t)}.
\end{align}
Since $m(i)-m'(i)\neq 0$ for at least one $i\in\{0,\ldots,t\}$, the absolute value of $n(m,m',t)$ is at least $1/M^t$. Thus by \eqref{eq:Omnull-diff},
\begin{equation}\label{eq:Omnull-lb}
	\lvert\hat x(m(0\!:\!t))-\hat x(m'(0\!:\!t))\rvert\geq\frac{\lvert\mc I_0\rvert}{M}\left(\frac{\lambda}{M}\right)^t.
\end{equation}
By Lemma \ref{thm:estdyn}, the right-hand side of \eqref{eq:Omnull-lb} equals $\ell_t$. Hence the lemma is proven.
\end{IEEEproof}

\begin{IEEEproof}[Proof of Lemma \ref{thm:mpdyn}]
	Recall the notation $\mc I=[\mc I_{\min},\mc I_{\max}]$ for real intervals $\mc I$. For $t\geq 0$,
	\begin{align}
		&\hat x(m(0\!:\!t+1))\\
		&\stackrel{(a)}{=}\mc I(m(0\!:\!t))_{\min}+\left(m(t+1)+\frac{1}{2}\right)\ell_{t+1}\notag\\
		&\stackrel{(b)}{=}\lambda\mc P(m(0\!:\!t))_{\min}-\frac{\Omega}{2}+\left(m(t+1)+\frac{1}{2}\right)\ell_{t+1}\notag\\
		&\stackrel{(c)}{=}\lambda\hat x(m(0\!:\!t))-\frac{\lambda\ell_t}{2}-\frac{\Omega}{2}+\left(m(t+1)+\frac{1}{2}\right)\ell_{t+1}\notag\\
		&\stackrel{(d)}{=}\lambda\hat x(m(0\!:\!t))\!-\!\frac{\lambda\ell_t}{2}\!-\!\frac{\Omega}{2}\!+\!\left(\!m(t+1)+\frac{1}{2}\right)\!\!\left(\frac{\lambda}{M}\ell_t+\frac{\Omega}{M}\right)\notag\\
		&=\lambda\hat x(m(0\!:\!t))+\frac{\lambda\ell_t+\Omega}{2}\left(\frac{2m(t+1)+1}{M}-1\right),\label{eq:hatxrec}
	\end{align}
	where $(a)$ is due to \eqref{eq:Pmn} and \eqref{eq:hatx}, $(b)$ is due to \eqref{eq:A(n),B(n)}, $(c)$ is again due to \eqref{eq:hatx} and $(d)$ is due to \eqref{eq:Pmnlength}. Therefore, 
	\begin{align}
		&\hat x(m(0\!:\!t+1))\notag\\
		&\!\stackrel{(e)}{=}\!\lambda\hat x(m(0\!:\!t))\!+\!\left(\!\frac{\lambda^{t+1}\lvert\mc I_0\rvert}{2M^{t+1}}\!-\!\frac{\lambda^{t+1}}{2M^t}\frac{\Omega}{M\!-\!\lambda}\!+\!\frac{\lambda}{2}\frac{\Omega}{M\!-\!\lambda}\!+\!\frac{\Omega}{2}\right)\!\times\notag\\
		&\qquad\qquad\qquad\qquad\times\left(\frac{2m(t\!+\!1)\!+\!1}{M}-1\right)\notag\\
		&=\lambda\hat x(m(0\!:\!t))+\frac{1}{2}\!\left(\frac{\lambda^{t+1}}{M^{t+1}}\lvert \mc I_0\rvert+\frac{\Omega M}{M-\lambda}\!\left(1-\frac{\lambda^{t+1}}{M^{t+1}}\right)\right)\times\notag\\
		&\qquad\qquad\qquad\qquad\times\left(\frac{2m(t+1)+1}{M}-1\right),\label{eq:step}
	\end{align}
	where $(e)$ is due to \eqref{eq:hatxrec} and \eqref{eq:ell-def}. Consequently, 
	\begin{align*}
	  &\hat x(m(0\!:\!t))\\
	  &\stackrel{(f)}{=}\lambda^t\biggl\{\hat x(m(0\!:\!0))\\
	  &\quad+\frac{1}{2}\sum_{i=0}^{t-1}\frac{1}{\lambda^{i+1}}\left(\frac{\lambda^{i+1}}{M^{i+1}}\lvert \mc I_0\rvert+\frac{\Omega M}{M-\lambda}\left(1-\frac{\lambda^{i+1}}{M^{i+1}}\right)\right)\times\\
	  &\qquad\qquad\qquad\times\left(\frac{2m(i+1)+1}{M}-1\right)\biggr\}\\
	  &\stackrel{(g)}{=}\lambda^t\biggl\{\hat x(m(0\!:\!-1))+\frac{\lvert\mc I_0\rvert}{2}\left(\frac{2m(0)+1}{M}-1\right)\\
	  &\quad+\frac{1}{2}\sum_{i=1}^{t}\!\left(\!\frac{\lvert\mc I_0\rvert}{M^i}\!+\!\frac{\Omega M}{M-\lambda}\!\left(\!\frac{1}{\lambda^i}\!-\!\frac{1}{M^i}\!\right)\!\right)\!\left(\!\frac{2m(i)+1}{M}\!-\!1\!\right)\!\biggr\}\\
	  &=\lambda^t\biggl\{\hat x(m(0\!:\!-1))\\
	  &\quad+\frac{1}{2}\sum_{i=0}^{t}\!\left(\!\frac{\Omega M}{M-\lambda}\!\left(\!\frac{1}{\lambda^i}\!-\!\frac{1}{M^i}\!\right)\!+\!\frac{\lvert\mc I_0\rvert}{M^i}\!\right)\!\left(\!\frac{2m(i)+1}{M}\!-\!1\!\right)\!\biggr\}.
	\end{align*}	
	where $(f)$ is due to Lemma \ref{lem:induction} and the recursion formula for $\hat x(m(0\!:\!t))$ derived in \eqref{eq:step} and in $(g)$ we applied \eqref{eq:Pmn} to find the relation between $\hat x(m(0\!:\!0))$ and $\hat x(m(0\!:\!-1))$. This completes the proof.
\end{IEEEproof}

\begin{IEEEproof}[Proof of Lemma \ref{thm:estandsecwithout}]
Without loss of generality, we may assume that $\hat x(m(0\!:\!T))>\hat x(m'(0\!:\!T))$. Then it is sufficient to show that if \eqref{eq:voraussohne} is satisfied, then $\hat x(m(0\!:\!T+t))-\hat x(m'(0\!:\!T+t))\geq\ell_{T+t}$ for all $t\geq 0$. We have
\begin{align}\label{eq:estdiff}
	&\hat x(m(0\!:\!T+t))-\hat x(m'(0\!:\!T+t))\notag\\
	&\stackrel{(a)}{=}\lambda^t\biggl\{\hat x(m(0\!:\!T))-\hat x(m'(0\!:\!T))\notag\\
	&\quad+\!\lambda^T\!\sum_{i=T+1}^{T+t}\!\left(\!\frac{\Omega}{M\!-\!\lambda}\!\left(\!\frac{1}{\lambda^i}\!-\!\frac{1}{M^i}\!\right)\!+\!\frac{\lvert\mc I_0\rvert}{M^{i+1}}\!\right)\!(m(i)\!-\!m'(i))\!\biggr\}\notag\\
	&\stackrel{(b)}{\geq}\lambda^t\biggl\{\hat x(m(0\!:\!T))-\hat x(m'(0\!:\!T))\notag\\
	&\quad-\!\lambda^T(M\!-\!1)\!\sum_{i=T+1}^{T+t}\left(\frac{\Omega}{M-\lambda}\left(\frac{1}{\lambda^i}-\frac{1}{M^i}\right)+\frac{\lvert\mc I_0\rvert}{M^{i+1}}\right)\biggr\}\notag\\
	&=\lambda^t\biggl\{\hat x(m(0\!:\!T))\!-\!\hat x(m'(0\!:\!T))\!-\frac{\Omega(M\!-\!1)}{(M\!-\!\lambda)(\lambda\!-\!1)}(1\!-\!\lambda^{-t})\notag\\
	&\quad-\frac{\lambda^T}{M^T}\left(\frac{\lvert\mc I_0\rvert}{M}-\frac{\Omega}{M-\lambda}\right)(1-M^{-t})\biggr\}
\end{align}
where $(a)$ is due to Lemma \ref{thm:mpdyn} and $(b)$ holds because $m(i)-m'(i)\geq -(M-1)$ for all $i$. Thus one obtains
\begin{align}
  &\frac{\hat x(m(0\!:\!T+t))-\hat x(m'(0\!:\!T+t))-\ell_{T+t}}{\lambda^t}\label{eq:goaldings}\\
  &\stackrel{(c)}{\geq}\hat x(m(0\!:\!T))-\hat x(m'(0\!:\!T))\notag\\
  &\quad-\!\frac{\Omega}{M\!-\!\lambda}\!\left(\!\!\frac{M\!-\!1}{\lambda\!-\!1}(1\!-\!\lambda^{-t})\!-\!\frac{\lambda^T}{M^T}(1\!-\!M^{-t})\!+\!\frac{1}{\lambda^t}\!-\!\frac{\lambda^T}{M^{T+t}}\!\right)\notag\\
  &\quad-\frac{\lvert\mc I_0\rvert}{M}\left(\frac{\lambda^T}{M^T}(1-M^{-t})+\frac{\lambda^T}{M^{T+t}}\right)\notag\\
  &=\hat x(m(0\!:\!T))-\hat x(m'(0\!:\!T))\notag\\
  &\quad-\frac{\Omega}{M-\lambda}\left(\frac{M-1}{\lambda-1}(1-\lambda^{-t})-\frac{\lambda^T}{M^T}+\frac{1}{\lambda^t}\right)
  -\frac{\lvert\mc I_0\rvert}{M}\frac{\lambda^T}{M^T}.\label{eq:dannauch}
\end{align}
where \eqref{eq:estdiff} and Lemma \ref{thm:estdyn} were used in $(c)$. Since we want \eqref{eq:goaldings} to be positive for every $t\geq 0$, it is sufficient by \eqref{eq:dannauch} to have
\begin{align*}
  &\hat x(m(0\!:\!T))-\hat x(m'(0\!:\!T))\\
  &\geq\max_{t\geq 0}\left\{\!\frac{\Omega}{M\!-\!\lambda}\!\left(\!\frac{M\!-\!1}{\lambda\!-\!1}(1\!-\!\lambda^{-t})\!-\!\frac{\lambda^T}{M^T}\!+\!\frac{1}{\lambda^t}\!\right)
  \!+\!\frac{\lvert\mc I_0\rvert}{M}\frac{\lambda^T}{M^T}\!\right\}\\
  &=\frac{\Omega}{M-\lambda}\left(\frac{M-1}{\lambda-1}-\frac{\lambda^T}{M^T}+1\right)
  +\frac{\lvert\mc I_0\rvert}{M}\frac{\lambda^T}{M^T}\\
  &\stackrel{(d)}{=}\frac{\Omega}{M-\lambda}\frac{M-1}{\lambda-1}+\ell_T,
\end{align*}
where $(d)$ is due to Lemma \ref{thm:estdyn}. Thus the inequality holds if \eqref{eq:voraussohne} is satisfied, which proves the lemma.
\end{IEEEproof}

\begin{IEEEproof}[Proof of Lemma \ref{lem:vol}]
	If we can show
\begin{align}
	&\hat x(m_{\xi(1:j-1)\xi(j)}(0\!:\!jT-1))-\hat x(m_{\xi(1:j-1)\xi'(j)}(0\!:\!jT-1))\notag\\
	&>\frac{\Omega}{M-\lambda}\frac{M-1}{\lambda-1}+\ell_{jT-1},\label{eq:bedvonanderswo}
\end{align}
for every $j\geq 1$, every $\xi(1\!:\!j-1)\in\{1,\ldots,\gamma\}^{j-1}$ and every $\xi(j),\xi'(j)\in\{1,\ldots,\gamma\}$ with $\xi(j)>\xi'(j)$, then the claim of the lemma follows from Lemma \ref{thm:estandsecwithout}. We have
\begin{align}
  &\hat x(m_{\xi(1:j-1)\xi(j)}(0\!:\!jT-1))-\hat x(m_{\xi(1:j-1)\xi'(j)}(0\!:\!jT-1))\notag\\
  &\!\stackrel{(a)}{=}\!\!\lambda^{jT-1}\!\!\!\!\!\sum_{i=(j-1)T}^{jT-1}\!\!\!\left(\!\frac{\Omega M}{M\!-\!\lambda}\!\!\left(\!\frac{1}{\lambda^i}\!-\!\frac{1}{M^i}\!\!\right)\!\!+\!\frac{\lvert\mc I_0\rvert}{M^i}\!\right)\!\frac{m_{\xi(j)}\!(i)\!-\!m_{\xi'(j)}\!(i)}{M}\notag\\
  &\!\stackrel{(b)}{\geq}\frac{\Omega}{M-\lambda}\frac{\lambda^T-1}{\lambda-1}\!+\!\left(\frac{\lvert\mc I_0\rvert}{M}\!-\!\frac{\Omega}{M-\lambda}\right)\!\frac{\lambda^{jT-1}}{M^{(j-1)T-1}}\frac{1-M^{-T}}{M-1}\notag\\
  &\!\stackrel{(c)}{=}\frac{\Omega}{M-\lambda}\frac{M-1}{\lambda-1}+\ell_{jT-1}+\frac{\Omega}{M-\lambda}\frac{\lambda^T-M-\lambda+1}{\lambda-1}\notag\\
  &\qquad+\left(\frac{\lvert\mc I_0\rvert}{M}-\frac{\Omega}{M-\lambda}\right)\left(\frac{\lambda}{M}\right)^{jT-1}\frac{M^T-M}{M-1}\notag\\
  &\!=:\frac{\Omega}{M-\lambda}\frac{M-1}{\lambda-1}+\ell_{jT-1}+A_{jT},\label{eq:daher}
\end{align}
where $(a)$ is due to Lemma \ref{thm:mpdyn}, $(b)$ uses $m_{\xi(j)}(i)-m_{\xi'(j)}(i)\geq 1$ which holds due to the choice of $\xi(j),\xi'(j)$, and Lemma \ref{thm:estdyn} was used in $(c)$. It remains to show that $A_{jT}\geq 0$. Since $\lambda^T\geq M+\lambda-1$ for $T$ satisfying \eqref{eq:T}, this is clear in the case that $\lvert\mc I_0\rvert/M\geq\Omega/(M-\lambda)$. Otherwise, we lower-bound $A_{jT}$ by $A_T$, for which we have
\begin{align*}
  &A_T+\frac{\Omega(M+\lambda-1)}{(M-\lambda)(\lambda-1)}\\
  &\geq\frac{\Omega}{M-\lambda}\lambda^T\left(\frac{1}{\lambda-1}-\frac{M}{\lambda(M-1)}\right)\\
  &\stackrel{(d)}{\geq}\frac{\Omega}{M-\lambda}\frac{\lambda(M-1)(M+\lambda-1)}{M-\lambda}\frac{M-\lambda}{\lambda(\lambda-1)(M-1)}\\
  &=\frac{\Omega(M+\lambda-1)}{(M-\lambda)(\lambda-1)}
\end{align*}
where $(d)$ is due to \eqref{eq:T}. This implies $A_T\geq 0$, hence $A_{jT}\geq 0$ for all $j\geq 1$. With \eqref{eq:daher}, this implies \eqref{eq:bedvonanderswo} for all choices of $j$, of $\xi(1\!:\!j-1)$ and of $\xi(j)>\xi'(j)$ and hence completes the proof of the lemma.
\end{IEEEproof}


\begin{IEEEbiography}[{\includegraphics[width=1in,height=1.25in,clip,keepaspectratio]{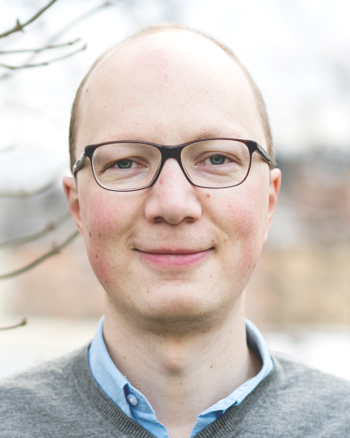}}]{Moritz Wiese}
(S’09-M’15) received the Dipl.-Math. degree in mathematics from the University of Bonn, Germany, in 2007. He obtained the PhD degree from Technische Universit\"at M\"unchen, Munich, Germany in 2013. From 2007 to 2010, he was a research assistant at Technische Universit\"at Berlin, Germany, and from 2010 until 2014 at Technische Universit\"at M\"unchen. Since 2014 he has been with the ACCESS Linnaeus Center at KTH Royal Institute of Technology, Stockholm, Sweden.
\end{IEEEbiography}

\begin{IEEEbiography}[{\includegraphics[width=1in,height=1.25in,clip,keepaspectratio]{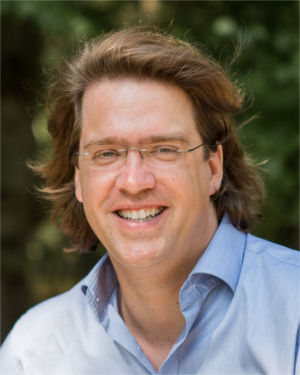}}]{Tobias J. Oechtering}
(S’01-M’08-SM’12) received his Dipl-Ing degree in Electrical Engineering and Information Technology in 2002 from RWTH Aachen University, Germany, his Dr-Ing degree in Electrical Engineering in 2007 from the Technische Universit\"at Berlin, Germany, and his Docent degree in Communication Theory in 2012 from KTH Royal Institute of Technology. In 2008 he joined the Communication Theory Lab at KTH Royal Institute of Technology, Stockholm, Sweden and has been an Associate Professor since May 2013. He is currently Associate Editor of IEEE Transactions on Information Forensic and Security since June 2016 and was editor for IEEE Communications Letters during 2012-2015. Dr. Oechtering received the ``F\"orderpreis 2009” from the Vodafone Foundation. His research interests include networked control, physical layer security and privacy, information theory, wireless communication as well as statistical signal processing.\end{IEEEbiography}

\begin{IEEEbiography}[{\includegraphics[width=1in,height=1.25in,clip,keepaspectratio]{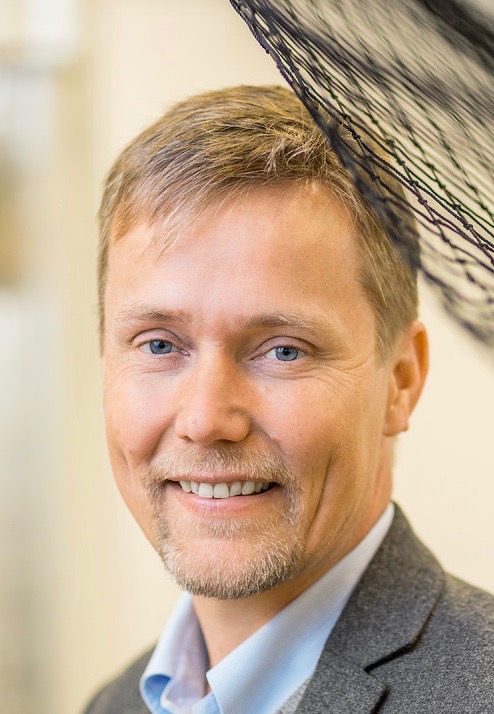}}]{Karl Henrik Johansson}
is Director of the Stockholm Strategic Research Area ICT The Next Generation and Professor at the School of Electrical Engineering, KTH Royal Institute of Technology. He received MSc and PhD degrees in Electrical Engineering from Lund University. He has held visiting positions at UC Berkeley, Caltech, NTU, HKUST Institute of Advanced Studies, and NTNU. His research interests are in networked control systems, cyber-physical systems, and applications in transportation, energy, and automation. He is a member of the IEEE Control Systems Society Board of Governors and the European Control Association Council. He has received several best paper awards and other distinctions, including a ten-year Wallenberg Scholar Grant, a Senior Researcher Position with the Swedish Research Council, and the Future Research Leader Award from the Swedish Foundation for Strategic Research. He is Fellow of the IEEE and IEEE Control Systems Society Distinguished Lecturer.
\end{IEEEbiography}

\begin{IEEEbiography}[{\includegraphics[width=1in,height=1.25in,clip,keepaspectratio]{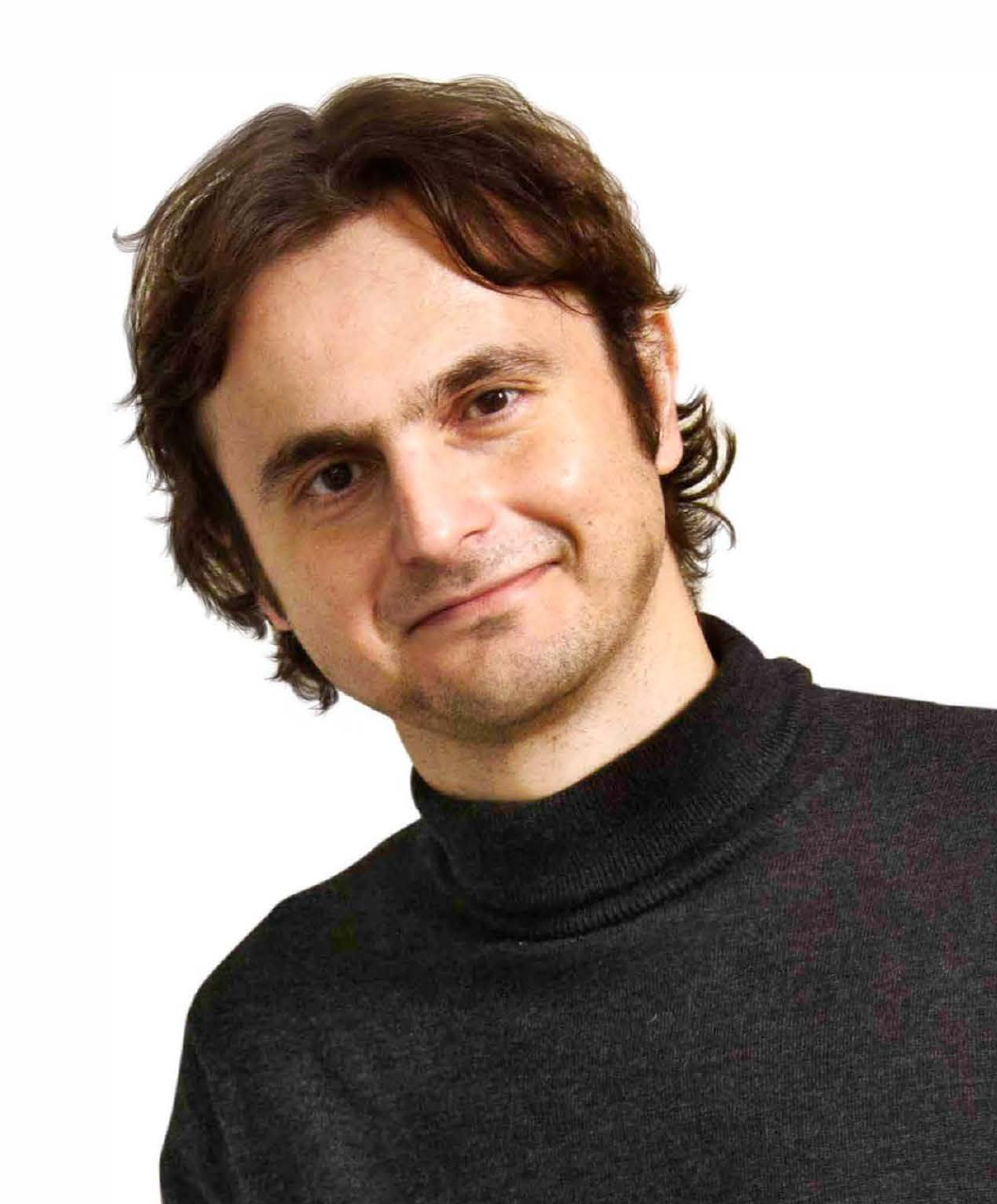}}]{Panagiotis (Panos) Papadimitratos}
earned his Ph.D. degree from Cornell
University, Ithaca, NY, in 2005. He then held positions at Virginia Tech,
EPFL and Politecnico of Torino. Panos is currently a tenured Professor at
KTH, Stockholm, Sweden, where he leads the Networked Systems Security
group. His research agenda includes a gamut of security and privacy
problems, with emphasis on wireless networks. At KTH, he is affiliated
with the ACCESS center, leading its Security, Privacy, and Trust thematic
area, as well as the ICES center, leading its Industrial Competence Group
on Security. Panos is a Knut and Alice Wallenberg Academy Fellow and he
received a Swedish Science Foundation Young Researcher Award. He has
delivered numerous invited talks, keynotes, and panel addresses, as well
as tutorials in flagship conferences. Panos currently serves as an
Associate Editor of the IEEE Transactions on Mobile Computing and the
ACM/IEEE Transactions on Networking. He has served in numerous program
committees, with leading roles in numerous occasions; recently, in 2016,
as the program co-chair for the ACM WiSec and the TRUST conferences; he
serves as the general chair of the ACM WISec (2018) and PETS (2019)
conferences. Panos is a member of the Young Academy of Europe.
\end{IEEEbiography}

\begin{IEEEbiography}[{\includegraphics[width=1in,height=1.25in,clip,keepaspectratio]{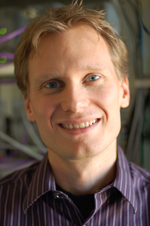}}]{Henrik Sandberg}
is Professor at the Department of Automatic Control, KTH Royal Institute of Technology, Stockholm, Sweden. He received the M.Sc. degree in engineering physics and the Ph.D. degree in automatic control from Lund University, Lund, Sweden, in 1999 and 2004, respectively. From 2005 to 2007, he was a Post-Doctoral Scholar at the California Institute of Technology, Pasadena, USA. In 2013, he was a visiting scholar at the Laboratory for Information and Decision Systems (LIDS) at MIT, Cambridge, USA. He has also held visiting appointments at the Australian National University and the University of Melbourne, Australia. His current research interests include security of cyberphysical systems, power systems, model reduction, and fundamental limitations in control. Dr. Sandberg was a recipient of the Best Student Paper Award from the IEEE Conference on Decision and Control in 2004 and an Ingvar Carlsson Award from the Swedish Foundation for Strategic Research in 2007. He is Associate Editor of the IFAC Journal Automatica and the IEEE Transactions on Automatic Control.
\end{IEEEbiography}

\begin{IEEEbiography}[{\includegraphics[width=1in,height=1.25in,clip,keepaspectratio]{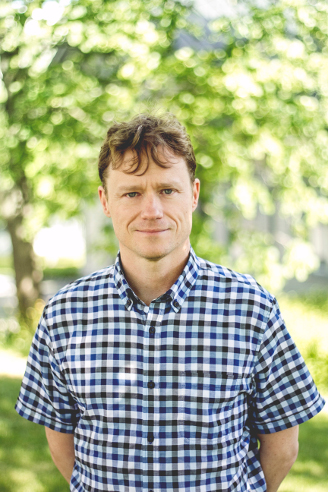}}]{Mikael Skoglund}
(S'93-M'97-SM'04) received the Ph.D.~degree in 1997
from Chalmers University of Technology, Sweden.  In 1997, he joined
the Royal Institute of Technology (KTH), Stockholm, Sweden, where he
was appointed to the Chair in Communication Theory in 2003.  At KTH,
he heads the Communication Theory Division and he is the Assistant
Dean for Electrical Engineering. He is also a founding faculty member
of the ACCESS Linnaeus Center and director for the Center Graduate
School.

Dr.~Skoglund has worked on problems in source-channel coding, coding
and transmission for wireless communications, communication and
control, Shannon theory and statistical signal processing. He has
authored and co-authored more than 130 journal and 300 conference
papers, and he holds six patents.

Dr.~Skoglund has served on numerous technical program committees for
IEEE sponsored conferences. During 2003--08 he was an associate editor
with the IEEE Transactions on Communications and during 2008--12 he
was on the editorial board for the IEEE Transactions on Information
Theory.
\end{IEEEbiography}

\end{document}